# On coherence and the transverse spatial extent of a neutron wave packet

C.F. Majkrzak, N.F. Berk, B.B. Maranville, J.A. Dura,
*Center for Neutron Research*,

and T. Jach
*Materials Measurement Laboratory*,

*National Institute of Standards and Technology*
*Gaithersburg, MD 20899*

(18 November 2019)

Abstract

In the analysis of neutron scattering measurements of condensed matter structure, it normally suffices to treat the incident and scattered neutron beams as if composed of incoherent distributions of plane waves with wavevectors of different magnitudes and directions which are taken to define an instrumental resolution. However, despite the wide-ranging applicability of this conventional treatment, there are cases in which the wave function of an individual neutron in the beam must be described more accurately by a spatially localized packet, in particular with respect to its transverse extent normal to its mean direction of propagation. One such case involves the creation of orbital angular momentum (OAM) states in a neutron via interaction with a material device of a given size. It is shown in the work reported here that there exist two distinct measures of coherence of special significance and utility for describing neutron beams in scattering studies of materials in general. One measure corresponds to the coherent superposition of basis functions and their wavevectors which constitute each individual neutron packet state function whereas the other measure can be associated with an incoherent distribution of mean wavevectors of the individual neutron packets in a beam. Both the distribution of the mean wavevectors of individual packets in the beam as well as the wavevector components of the superposition of basis functions within an individual packet can contribute to the conventional notion of instrumental resolution. However, it is the transverse spatial extent of packet wavefronts alone -- over which the phase is sufficiently uniform -- that determines the area within which a coherent scattering process can occur in the first place. This picture is shown to be entirely consistent in principle with the formal tenets of standard quantum theory. It is also demonstrated that these two measures of coherence can be distinguished from one another experimentally in practice.

Introduction

It is has become well-established over many years of research that for the analysis of most neutron scattering data in studies of condensed matter structure, it suffices to treat the incident and scattered neutron beams as if composed of distributions of plane waves with wavevectors of different magnitudes and directions. These distributions (typically assumed to be Gaussian) are conventionally taken to correspond to an instrumental resolution which effectively limits the degree to which neighboring features in a pattern of diffracted intensity, as a function of scattering angle, can be distinguished (e.g., [1] and references cited therein). Although this representation of the wavefunction associated with a freely propagating neutron quantum particle in a beam as a single, completely unlocalized plane wave state may be a useful mathematical idealization for the analysis of a broad range of diffraction measurements, it is, nonetheless, in principle, strictly unphysical. Moreover, a description of neutron scattering based on the aforementioned plane wave approximation may not be completely adequate even in practice in certain situations -- such as where the actual, finite spatial extent of a neutron wavefront -- a quantity intrinsic to a more realistic characterization of an individual neutron wavefunction as a spatially localized packet -- is of importance. A relatively common example involves an inhomogeneous material consisting of a mixture of different component volumes or domains, each producing a distinct scattering response. The transverse spatial dimensions of an individual neutron packet (i.e., perpendicular to its mean direction of propagation) then determine whether an interaction can occur with either one domain at a time or several neighboring domains simultaneously. (The coherent spatial extent of a wave packet parallel to the direction of the mean wave vector, the longitudinal component, is neither directly accessible in the measurements considered in this work nor is it typically of great significance in elastic scattering processes as long as it is sufficiently greater than the transverse length.) In the former case, where the packet wavefront is sufficiently smaller than the size of any one material domain, the net reflected beam intensity observed would simply consist of an incoherent (i.e., non-interfering) sum of *intensities* contributed by each of the various domains in the material illuminated by the incident beam. But for the latter possibility, in which the packet wavefront is of greater size than a number of neighboring material domain volumes altogether, the reflected wave packets which would be observed would each be composed of a coherent superposition of component wave *amplitudes* from every one of the contributing material domains, thereby exhibiting interference effects. Clearly, knowing which of these two limiting possibilities (or an intermediate combination thereof) applies is necessary if an accurate analysis of the scattering data is to be performed. Another specific instance in which the spatial size of the neutron wave packet matters pertains to the formation of orbital angular momentum (OAM) states of the neutron. The transverse extent over which the phase of a neutron packet wavefront is of the requisite uniformity determines whether or not a suitably structured material object of a given size can impart orbital angular momentum (OAM) to a neutron [2].

Consequently, instead of the approximate, conventional picture of instrumental resolution involving incoherent plane wave distributions, in certain circumstances an instrument beam needs to be more accurately described as a collection of independent and non-interacting individual neutron quantum particles each of which has an associated, localized packet wavefunction. Slow neutrons, having wavelengths between approximately 1 and 10 Angstroms, as typically employed in scattering studies of condensed matter structure, interact with material as a wave, but do so one individual neutron at a time -- in a manner first emphasized by Dirac [3]. (For light scattering, Maxwell's equations for both electric and magnetic fields need to be solved, requiring the introduction of many-body photon (Boson) states to describe the light produced, for example, by a laser source -- unlike the case for neutrons which, as Fermions, do not form such states.) Relevant characteristics of wave packet functions are considered in the following section "*The Free Neutron -- A Quantum Particle with an Associated Wave Packet*" and "*Appendix A: Gaussian Wave Packets*". Moreover, a *beam* can be quantitatively described in terms of the coherent superposition of basis state functions and corresponding wavevectors composing each individual packet in conjunction with an incoherent (non-interfering) ensemble of packets with corresponding distributions of the magnitudes and directions of constituent packet mean wavevectors. A schematic representing such a view of a neutron beam appears in Figure 1 [4]. This interpretation, as presented herein, is neither implied to be entirely unique nor original -- although we are not aware of an explicit, comprehensive exposition elsewhere in the literature, at least in the present context. In any event, we also present here new evidence to support such a picture which should contribute to the overall state of knowledge on the subject that might be of interest not only for neutron scattering but other applications involving Fermions, particularly electrons, as probes of condensed matter.

Both the collective and individual distributions contribute to the description of the instrumental resolution insofar as the ability to distinguish features in a diffraction pattern is concerned. On the other hand, it is the coherent superposition of basis states which compose an individual packet wavefunction alone that determines the transverse (i.e., perpendicular to the mean packet wavevector) extent over which a particular wavefront is of sufficiently uniform phase to interact simultaneously with scattering material in a coherent manner (forming a superposition of reflection amplitudes).

For the reasons stated earlier, it can be useful in the analysis of scattering data to distinguish between two different, though related, measures of coherence -- one pertaining directly to a collective beam resolution and the other to the transverse extent of an individual neutron packet wavefront. However, to do so in practice can be problematical since both the collective characteristics of the beam and the individual properties of each constituent packet are defined to some degree by the very same instrumental devices. These include collimators, such as a pair of slit apertures in series to define angular divergence, and single crystal monochromators to determine (primarily) the wavelength spread through Bragg diffraction from a set of parallel atomic planes. Reflecting mirrors and refracting prisms, their action described in a manner similar to that used in ordinary light optics, are also commonly employed devices which can shape both packet and beam.

The section of the paper entitled "*Diffraction from a Single Slit Aperture*" summarizes how the elementary process of diffraction from a slit aperture can shape the wavefront of the transmitted neutron packet. It is shown that for typical instrumental parameters, the wavefronts so created may not always be described in the far-field or Frauhofer limit. The description of a beam and its individual constituent individual neutron packets emerging from a pair of slits in series -- illuminated by a crystal monochromator -- is then considered in the subsequent sections "*Angular Divergence Defined by a Pair of Slits in Series*" and "*Appendix B: Reflection from Perfect Crystal Mosaic Blocks*". Measurements of the profile of a beam formed in such a manner are shown to be in excellent agreement with this model of a neutron beam composed of independent wavepackets. And as further demonstrated in the section that immediately follows, "*Diffraction from Phase Gratings in Near-Normal Transmission*", the diffraction pattern observed by transmitting the same crystal-monochromated / slit-collimated beam through a phase grating is also well-described by the ensemble of packets model for a beam.

Because of relatively recent applications of a theory originally introduced to describe partially coherent and quasi-monochromatic ordinary light from temporally and spatially extended incoherent sources (e.g., [5]; [6]; [7]) to both neutron ([8], [9], [10]) and x-rays ([11], [12], [13], [14] -- see also the more recent article by J. Stoehr [15] and references therein), it is also of relevance to re-examine its connection to the meaning of neutron coherence. In particular, it is shown in the section "*Incoherent Beams: The Mutual Coherence Function*" that it is the incoherent distribution of packet mean wavevector directions (or, equivalently, the beam angular divergence defined primarily by the geometry of collimating devices such as a pair of slit apertures in series) which can be more closely related to the concepts of fringe visibility and related mutual coherence functions. This assertion follows when the mutual coherence function and connected relationships are derived in the Fraunhofer or far-field limit, assuming plane wave functions -- as is typically done.  Under these specific conditions, the mutual coherence function is effectively a measure of the extent to which a spatially extended incoherent source reduces the resolving power of a perfect plane wave (approximated by a single coherent point source an infinite distance away).

One way to differentiate beam and packet contributions to the resolving power of an instrument and to determine the transverse extent over which the phase of a neutron packet wavefront is uniform to within a specified tolerance is considered in the section "*Coherent Averaging by a Wave Packet*".  Normally, in scattering studies of ordered systems, knowing the resolution of the instrument establishes limits on the scale over which the correlations between structural features in that material can be assessed.  But from another point of view, a known correlation in a periodic sample can be used as a measuring tool to infer the transverse coherent extent of a neutron packet wavefront [4].  Nonetheless, determining this transverse extent from line-width analysis can sometimes be problematical because of the overlapping coherent and incoherent contributions to the broadening of observed diffraction patterns.  Alternatively, an analysis of the total external reflection below the critical angle at glancing angles of incidence from thin film gratings can reveal the transverse coherent extent of a neutron packet wavefront almost independently of beam angular divergence.  New measurements employing gratings in this way are reported in the present work -- in the section "*Reflection from Grating Structures at Glancing Angles*" -- which further quantify such methods.  Nonetheless, the aforementioned techniques are subject to the limitations imposed by non-ideal reference objects -- such as grating structures deposited on substrates that are not perfectly flat -- as described in the section "*Like the Wavy Surface of a Circus Mirror*".

In the final section of this paper, "*Description of a Neutron Wave Packet According to Standard Quantum Theory*", it is shown that the picture of a beam composed of independent neutrons, each represented by a distinct quantum particle together with corresponding wave packet, as presented herein, is, in general, rigorously consistent with standard quantum theory.  It is further demonstrated that the interpretation of density operators -- for the pure and mixed states corresponding to individual neutron packets and beams, respectively -- as recently put forward by Berk [16] is in agreement with observation.  This discussion specifically addresses a possible misconception that emerged in the past concerning the meaning of density operators as applied to the description of neutron beams.

The Free Neutron -- A Quantum Particle with an Associated Wave Packet

Consider again the typical experimental configuration schematically represented in Figure 1.  Imagine that a single neutron is emitted via a fission reaction from its origin within the nucleus of a uranium atom as a freely propagating quantum particle.  Its energy is subsequently moderated by inelastic collisions with the nuclei of molecules such as heavy water or liquid hydrogen within a finite volume which effectively acts as a temporally and spatially extended incoherent source.  The confinement of a neutron to the space between source and detector, bounded by fixed instrumental components such as guide tubes and apertures, requires that its state be defined as a solution of the Schoedinger equation having limited spatial extent -- i.e., a state traditionally referred to as a wave packet.  Every individual wave packet can be associated with a mean wavevector.  If that mean wavevector is within a range of magnitudes and directions accepted by a monochromating device and set of apertures, the neutron then has the possibility of being incident upon and scattered by a material sample object into a detection device -- wherein it may be captured to again become a bound constituent of another nucleus.

One possible mathematical representation of a free neutron wave packet that is widely adopted is a function $\Psi(\mathbf{r},t)$ made up of a weighted distribution of plane wave momentum eigenstates (basis) which may be written as

(1) $$\Psi(\mathbf{r},t) = \int A(\mathbf{k}) \exp[i(\mathbf{k} \cdot \mathbf{r} - \omega_k t)]\, d\mathbf{k}$$

where $\mathbf{r}$ is a spatial coordinate, $\mathbf{k}$ is a basis state wavevector, $\omega_k$ is a corresponding angular frequency, t is the time, and $A(\mathbf{k})$ describes a distribution of the basis states. This real space representation and its momentum space counterpart are, as is well known, related by a Fourier transform. The effective size and shape (in real space) of an individual neutron wave packet -- including the extent over which a given wavefront is at a uniform phase -- is related to the coherent superposition of momentum basis states forming the packet through that Fourier transform. This connection can be expressed (famously) as an uncertainty relation which applies along each Cartesian direction, including the orthogonal directions perpendicular (transverse) and parallel (longitudinal) to the mean packet wavevector. Intrinsic uncertainties in position and momentum, corresponding to the widths of the distributions in momentum $\Delta k$ and coordinate $\Delta r$ for each rectangular component are related -- for the case of (minimum uncertainty) Gaussian wave packets -- by (in one dimension)

(2) $$\Delta k_X\, \Delta x \geq 1/2$$

(see Appendix A for further relevant details). Whereas a single plane wave basis state is a solution of both the time-dependent and time-independent Schroedinger equations of motion, the 3D wave packet of Equation 1 satisfies only the explicitly time-dependent version. This has consequences regarding physical interpretations as will be discussed subsequently. The time-independent wave equations for neutrons and x-rays with their corresponding stationary state solutions as well as the three-dimensional, free-particle Schrödinger wave packet that is a solution only of the time-dependent equation of motion (and almost universally described by the "standard" Gaussian model) are discussed in Appendix A.

Diffraction from a Single Slit Aperture

Short of describing the neutron wave function as a more realistic -- but far more complicated -- wave packet localized in all three dimensions, it is possible to obtain a wave function that is partially localized in the two orthogonal transverse directions (perpendicular to the mean propagation wavevector along the x-axis in the figure) but which is also a stationary state solution of the time-*independent* Schroedinger equation. This can be achieved by a suitable superposition of wavevector *directions* about the x-axis (i.e., a distribution of y- and z- components) such that the magnitudes k of all the component wavevectors are equal -- that is, $k^2 = k^2_X + k^2_Y + k^2_Z$ = a constant value. For appropriate distributions, say Gaussian, of y- and z-components of the wavevector (with the propagation direction along the x-axis), the resultant waveform resembles a tubular-like wave localized about the x- and-z-axes but extended along the x-axis. This particular form is suitable for several derivations that appear in subsequent sections since, importantly, the essential characteristics of the phenomena of interest are preserved.  This simplified waveform is similar to the so-called "quasi-monochromatic wavetrain" solution of the approximate paraxial wave equation employed in elementary treatments of the diffraction of light (e.g., [7]).  In a typical example, an elongated packet waveform of this type for an individual neutron might have a distribution of plane wave basis states corresponding to an angular range of wavevector orientations of the order of $\epsilon$ = 5 arc seconds (2.424 x $10^{-5}$ radians) about the mean wavevector direction.  In this case, the constraint that $k^2$ be a constant value (for, say, k = 2 $\pi$ / 5 Å) results in a magnitude variation of the longitudinal wavevector component of approximately k [ 1 - cos ($\epsilon$) ] = 2.94 x $10^{-10}$ k whereas the transverse (perpendicular) component variation is k sin( $\epsilon$ ) =  4.23 x $10^{-7}$ k (or 1439 times larger).  Another way of expressing this approximation is to require that the magnitude of the width of a distribution of transverse wavevector components $\Delta k_\perp$ << |$\mathbf{k}$| where $\mathbf{k}$ is along the longitudinal direction.  This approximation has been adopted in the description of initial electron wave functions where orbital angular momentum is consecutively imparted by an appropriate device [17].  For the purposes at hand then, it suffices to represent neutron wave functions as nearly plane wave fronts of truncated lateral extent produced, for example, by perfect plane waves incident on and diffracted from a single slit -- as will be demonstrated below.

Consider once again Figure 1 and focus on the second, downstream slit of width W.   Assume that, prior to interaction, the wave function of a freely propagating neutron incident on this slit from the left can be well-enough represented by a single plane wave of wavelength $\lambda$ with wavevector parallel to the x-axis as depicted in Figure 2.  This corresponds, in an idealized mathematical limit, to a neutron emitted from a single point source an infinite distance away on the negative x-axis.  In such a case, each wavefront has sufficient lateral extent to span the entire aperture width W in phase -- which significantly simplifies the subsequent mathematical description of the resulting diffraction pattern but without sacrificing its essential characteristics of interest here.   In the two-dimensional system depicted in Figure 2, the aperture is taken to be defined by a pair of totally absorbing masks.  The resultant diffraction pattern of scattered intensity I on a line of observation a distance S from the slit far enough away to be in the far-field or Fraunhofer limit (S > $W^2$ / $\lambda$.) is given by the well-known expression (e.g., [7])

(3)                                    $I(\theta) = I(0) (\sin \beta / \beta)^2$

where $\beta = (\pi W / \lambda) \sin \theta$ and $\theta$ is the angle from the bisecting perpendicular to the aperture at the center of the opening to a line connecting the slit center to a point on a detecting plane (I(0) represents the incident wave intensity).  The square of the sinc function on the RHS of Equation 3 has a central maximum with a FWHM $\Delta\theta_{SSD}$ (subscript SSD indicates single-slit diffraction) *approximately* equal to the first zero of sinc $\beta$  which is given by $\theta = \arcsin(\lambda/W)$.

Neglecting higher-order multiple scattering effects, a simple Huygens-Fresnel wavelet construction can further provide a relatively accurate picture of both the amplitude and intensity distribution of the waveform emanating from this single aperture over a significant range of distances from the near-field Fresnel region out to the far-field Fraunhofer limit.  The results of such a Huygens-Fresnel construction for a variety of pertinent aperture widths and distances to a plane of observation are represented in Table 1 and Figure 3 for the two-dimensional geometry shown in Figure 2.  (The diffraction of neutrons by slit widths of the order of 100 microns was first conclusively demonstrated by Shull et al. [18].)

One measure of the distance over which a given wavefront is uniform in phase to within one wavelength can be determined by examining two consecutive wavefronts propagating along the x-axis as pictured in the right-hand column of Figure 3 as follows. Identify the point of maximum amplitude on the ridge of the leading wavefront (shown as a blue curve) that has the same transverse y-value as that of a point on the ridge of the following front (the intersection of the blue curve and red line). The x-coordinates of these two points differ by λ. In Figure 3, for the 10 micron aperture width in the far-field limit, this measure corresponds roughly to that obtained from the first minimum of the intensity on either side of the central maximum. Considering the amplitude waveforms on the right of the figure it is clear that the widths of the wavefronts propagating outward from the aperture posses a finite lateral extent, at least over which the probability amplitude is of significant magnitude and of uniform phase within one wavelength.

Thus, even if the neutron wave function approaching the slit is represented by a single plane wave, the interaction with any aperture of finite size transforms the wave function into one that is localized in space to a finite extent in a transverse direction. And although the transverse width continues to increase with distance from the aperture, as described by the Fraunhofer diffraction formula, it remains of finite size at a finite distance away from the aperture.

It should be emphasized that unless the incident wavefront spans the width of a given aperture, the standard single-slit diffraction pattern will not be formed, although in some cases diffraction from one of the edges of the masks defining the aperture may occur. But in general, if the wavefront has a transverse extent less than the width of the aperture, the aperture acts primarily in the geometrical optics limit to define the spatial and/or angular range over which rays emanating from an upstream source (be it a point or extended) can pass through. A pair of slits in series, for example, can also define the angular divergence of a beam of individual neutrons, each with its own corresponding packet. Thus, the allowed angular range of directions for the mean wave vector of each packet can be limited by the geometrical angular range defined by the pair of slits in series.

Angular Divergence Defined by a Pair of Slits in Series

Consider next a case in which a beam of neutrons, each individual neutron still associated with a wave train of finite transverse dimensions as described in the preceding section, but where the mean wavevector direction of each wave train (2D elongated packet) can vary over a range of angles centered about the average direction of the beam -- as might be produced by the arrangement shown in Figure 1. The monochromator device as it is drawn is intended to represent a mosaic crystal composed of an angular distribution of perfect single crystalline blocks of finite dimensions (e.g., pyrolytic graphite) which selects and re-directs (via Bragg diffraction) a fraction of the neutrons incident upon it through the pair of downstream apertures. Imagine, for the time being, the limiting case where the transverse spatial extent of an individual incident neutron's wave packet is sufficiently small that it can only interact with one perfect single crystal mosaic block at a time. Each mosaic block can then be treated as a miniature monochromator and collimation device which -- by the coherent Bragg reflection process -- helps to define the shape (form) and extent (width) of the transverse (perpendicular to the mean packet wavevector) and longitudinal (parallel to the mean wavevector) distributions of wavevectors corresponding to the set of basis wave functions that compose an individual neutron wave packet. In this case, the mosaic blocks, which are not necessarily all of exactly the same size nor are uniformly spaced from one another, are randomly oriented according to a Gaussian distribution of angles about a mean normal direction. Although each separate block is a coherent reflector of single neutrons, collectively, the ensemble of blocks together can be considered to be a spatially incoherent secondary source of a beam of neutrons. This distribution of mosaic blocks is analogous to a spatially incoherent distribution of point sources. But in contrast to the spherical waves that emanate from point sources, isotropically, the blocks radiate over a limited range of preferred directions and with a quasi-monochromatic bandwidth. The reflected neutron wave amplitude from a given block is a result of the constructive interference that occurs because of the periodic structure of a number of parallel atomic planes in that block. The size of the block affects the distribution of basis state functions composing the reflected neutron wave packet and thereby the area over which a component wavefront of the packet is of uniform phase. At sufficiently large distances, in the far-field or Fraunhofer limit, and under the proper initial conditions, the packet wavefronts generated by a mosaic block are comparable to those of truncated plane waves or the quasi-monochromatic wave trains as discussed in the preceding section. A more detailed discussion of diffraction from a perfect single crystal mosaic block can be found in Appendix B below.

The pair of apertures can also act in unison to define the distribution of wave packet directions within a beam. If the aperture widths are sufficiently large that diffraction effects are negligible, then the primary effect of the apertures is to define a collective beam contribution to angular divergence with an approximate FWHM $\alpha$ where tan $\alpha$ = W/L.

As an example, for W =1 mm, the diffraction width predicted by the far-field Fraunhofer expression of Equation 3 for $\lambda$ = 5 Å is $\Delta\theta_{SSD}$ = 2.865 x $10^{-5}$ deg whereas a 0.1 mm slit width would give 2.865 x $10^{-4}$ deg. The corresponding geometrical angular widths $\Delta\alpha$ = arctan(W/L) as defined by a pair of apertures of the same width W a typical distance L = 1500 mm apart are 3.82 x $10^{-2}$ and 3.82 x $10^{-3}$ deg -- thus the fractional increase in angular divergence caused by diffraction over that due to the geometrical width is only approximately 0.075 % and 7.5 %, respectively.

Note that there is no unique combination of W and L which define a particular geometrical angular divergence $\Delta\alpha$ -- in the absence of appreciable diffraction broadening from the slits, this geometrical angular divergence is constant with increasing distance downstream at any point of observation. It can only be meaningfully associated with the beam divergence arising from a distribution of mean wavevectors of the individual neutron packets composing a beam.

Given the physical description of the instrumental components that define the beam and individual neutron wave functions presented above, an accurate representation of the beam profile for an actual instrument can be calculated. Further detailed specifications of the instrument on which the measurements to be described in following sections of this work are given in Appendix D. In addition, the action of the crystal monochromator, in particular, is discussed at length in Appendix C. It is expected that for typical HOPG monochromators the reflected neutrons have wave functions with transverse dimensions of the order of tens of microns at a distance of a meter or so away (this is consistent with the diffraction measurements described below and in other works, e.g., [19]).

Based on the Huygens-Fresnel numerical calculation discussed in the previous section, diffraction by a slit of width 0.025 mm for 5 Å wavelength neutrons produces a wavetrain with a transverse dimension of the same order as the slit width at a distance of one meter away, assuming of course that the wavefront of the incident packet was of sufficient planarity and width to span the aperture in the first place. The diffraction broadening by such a slit is only several arc seconds (~ 3.44 arc secs). This is negligible compared to the angular divergence (more than one degree) defined by the width of the monochromator source (approximately 25. mm) which illuminates the first downstream slit roughly a meter away. Thus, we can, to a good approximation, view the first (upstream) slit of the collimating pair of slits in series as a uniform (in space and angle) source which subsequently illuminates the second (downstream) slit. However, given the widths and distance between the slits, the geometrical angular divergence of a beam defined by and emanating from the pair is $2.89 \times 10^{-5}$ radians (1.66 degrees = 0.0995 minutes = 5.97 arc seconds). Thus the beam defined by the pair of slits together has a geometrical angular divergence that is comparable (roughly twice as large) to the diffraction broadening associated with the downstream slit.

For the specific values described above, a numerical calculation of the beam profile expected to be projected onto the detector line on the instrument can be performed and compared to an actual measurement as shown in Figure 4. The computed curve is not a fit, but only scaled to the measured intensity. A slightly larger slit width of 0.030 mm was found to be in better agreement than the nominal value of 0.025 mm of the shim stock spacer used to define the gap. This is likely due to non-perfect alignment of the slits with respect to the vertical axis and imperfections in the machined edges of the 1 mm thick Cd masks, and possibly also due in part to mirror reflection and refraction as well as diffraction from the mask edges by packet wavefronts of insufficient lateral extent to span both mask edges simultaneously. In this calculation, the geometrical angular limits defined by the pair of slits together were taken into account and intensities -- as given by the standard Fraunhofer diffraction formula -- from source points across the width of the first (upstream) slit were summed. In Figure 4, the measured data are compared to the results of two calculations, one corresponding to that expected for geometrical ray optics alone and the other to a combined result due to both geometrical and diffraction effects.

Diffraction from Phase Gratings in Near-Normal Transmission

Let us consider next a beam made up of neutrons with the same 2D longitudinally elongated packet wave functions described above, illuminating a phase grating at normal incidence (in transmission geometry). For neutrons with a 5 Å nominal wavelength, single crystal silicon has a scattering length density ($2.1 \times 10^{-6}$ Å$^{-2}$) which produces a $\pi$ phase shift over a distance of about 30 microns. By etching a parallel set of grooves of uniform width and spacing and of rectangular cross section into a plate of perfect single crystal silicon, a transmission phase grating can be fabricated. The number of grating periods which simultaneously interact (i.e., coherently) with an individual neutron incident upon the grating (perpendicular to the surface in which the grooves were etched) depends on the transverse extent of a neutron packet wavefront over which the phase is sufficiently uniform. For example, if the wavefront uniformly spans a width on a $\pi$ phase-shift grating equivalent to N periods, the diffracted intensity $I_{PG}$ in the far-field or Fraunhofer regime depends upon N directly as given by

(4)        $I_{PG} = 2\ I_0\ [\ \sin(\ \beta\ )\ /\ \beta\ ]^2\ [\ \sin(\ N\alpha\ )\ /\ \sin(\ \alpha\ )\ ]^2\ [\ 1 - \cos(\ \alpha\ )\ ]$

where

        $\alpha = (ka/2)\sin(\ \theta\ )$   and   $\beta = (kb/2)\sin(\ \theta\ )$

in which b is the groove width, a is the grating spacing or period, $I_0$ is the incident intensity and $k = 2\pi / \lambda$ (where k and $\lambda$ are the nominal neutron wavevector and wavelength, respectively). Notably, a plot of $I_{PG}$ versus $\theta$ reveals that the centers of the principal maxima about the origin are markedly shifted for N =1 compared to the positions for N $\geq$ 1, as will be shown below. The angle $\theta$ is defined similarly as the angle of diffraction shown in Figure 2 for the case of the single slit. Note that Equation 4 is explicitly for the case where the neutron wavevector is exactly perpendicular to the plane of the grating. For a beam composed of packets with a distribution of mean wavevector directions, i.e., with a geometrical angular divergence, the intensity contributions, properly weighted according to the particular distribution (e.g., Gaussian), must be summed over the range of incident angles in that distribution. For relatively narrow angular distributions, the center of the diffraction pattern corresponding to a given incident angle is, in the small angle approximation, effectively shifted from that for normal incidence by an amount approximately equal to the difference in the given angle of incidence from the normal.

Figure 5 shows diffraction patterns calculated for a model $\pi$ phase-shift grating with equal column and trough thicknesses at a 5 Å neutron wavelength. One set of patterns corresponds to coherent contributions from 2 grating periods and the other set from 4. The period of the grating is 2.4 microns. Assuming the grating period itself is perfectly uniform, the number of coherently contributing periods then depends on the transverse width of the neutron wavefront over which the phase is of the requisite uniformity (e.g., a truncated plane wavefront). The angular divergence of the incident neutron beam -- which corresponds to a distribution of transverse components of packet mean wavevectors -- determines how well the features of the pattern are resolved. Both figures include cases for zero beam divergence and two other finite values (3.6 and 7.2 sec of arc) convoluted with the natural pattern. Note that the general shape of the pattern is preserved while the widths and magnitudes of the principle and subsidiary reflections are affected by the breadth of the geometrical angular divergence of the incident beam.

Table 2 compares the width of the transverse wavevector component distribution $\Delta k_{T\,WP}$ associated with an individual wave packet -- possessing a corresponding spatial transverse width $\Delta r_{T\,WP}$ -- with that of the packet mean wavevectors $\Delta k_{MT(BEAM)}$ contained within the beam.

Figure 6 shows a diffraction pattern measured on the MAGIK reflectometer at the NCNR from an 8 micron period $\pi$ phase-shift grating with equal-thickness rectangular troughs and columns etched in single crystal silicon [20] at a neutron nominal wavelength of 5 Å. Also plotted in this figure is a calculated model diffraction pattern based on the phase grating formula of Equation 4 and assuming an incident illuminating beam described exactly as in the preceding section for the pair of slits which resulted in the profile of Figure 4 (the phase grating was located 495. mm away from the second downstream slit). That is, both geometrical and diffraction effects in forming the beam incident on the grating by the pair of slits were taken into account. The phase grating diffraction pattern was subsequently computed assuming both the distribution of geometrical angles and the transverse dimension of an individual neutron packet wavefront (over which the phase was uniform).

The model calculation was not fit to the data, but only scaled to the measured intensity (as was done previously for the pair of slits). The best agreement between the data and model was obtained for N = 3 (corresponding to a distance of 24. microns) and for a slight curvature of the grating substrate amounting to about 2.65 x $10^{-5}$ radians (5.47 arc seconds).  (The bending might alternatively be attributed to a curvature of a neutron packet wavefront -- which was originally taken to be perfectly flat but limited to a 24. micron finite lateral extent.  This wavefront dimension is consistent with the width of the slit.)

In related work by Treimer et al. [19], diffraction patterns were measured for single slits of various widths (including 100 microns) as well as for multiple-groove gratings (periods of 16 and 32 microns) with neutrons prepared on a USANS instrument (nominal wavelength of 5.248 Angstroms using an HOPG (002) pre-monochromator and a pair of 7-bounce channel cut Si (111) crystals as monochromator and analyser ).  The definition of the beam angular divergence by channel cut Si crystals instead of a simple pair of slits results in a significantly cleaner beam, free from spurious artifacts potentially caused by edge effects of the masks used in our instrumental set-up discussed above.  In their grating diffraction experiments, measurements were performed at two different values of the incident beam angular divergence -- 1.4 and 5.7 seconds of arc FWHM.  The diffraction pattern features were found to be better resolved at the tighter angular resolution of the beam, as would be expected based on the description of a beam of neutrons which we have presented above.  Moreover, it was also observed that the diffraction patterns obtained in their experiments could be fit to a high degree of accuracy assuming a transverse extent of the neutron packet wavefront of 80 microns FWHM, *irrespective* of whether the angular spread of the beam was 1.4 seconds of arc (2.33 x $10^{-2}$ minutes of arc = 3.89 x $10^{-4}$ degrees = 6.79 x $10^{-6}$ radians) or about 4 times larger at 5.7 arc-seconds (2.76 x $10^{-5}$ radians).  *If* the distribution in the transverse wavevector components of an individual neutron wave packet were to be attributed to these beam geometrical angular divergences $\Delta\alpha$, a transverse wavevector distribution subsequently computed from $\Delta k_T = k \Delta\alpha$ would predict transverse wave packet widths only of the order of  $\Delta r_T = $ 6.15 and 1.51 microns, respectively, via the uncertainty relation $\Delta k_T \Delta r_T = 1/2$ (Equation 2).  This would clearly be at odds with the experimental finding of 80 microns for the transverse extent of the neutron packet wavefront.

In summary, the evidence presented thus far strongly suggests the existence of two distinct distributions -- one being that of the mean wavevector values of all of the neutron packets comprising the incident beam and the other a distribution of transverse wavevector components associated with the coherent superposition of momentum basis states that form a single neutron packet wave function.  In other words, these findings are consistent with the picture of a neutron beam composed of an ensemble of similar, individual neutron quantum particles each associated with a packet wave function that represents a pure state.

Incoherent Beams: The Mutual Coherence Function

In this section, we focus on how a radiation source affects the instrumental beam resolution.  Consider a number of different possible extended (line) source types to be represented by multiple apertures, as shown  schematically in Figure 7.  The coherence of the radiation emitted from such a secondary source depends upon the nature of the radiation incident on the aperture array from a primary source to the left.  A number of possible cases are depicted with corresponding descriptions on the left-hand side of the figure.  For the discussion to follow we will assume case (B) in which the line source is completely spatially and temporally incoherent -- and also assume that the source is a continuous emitter of circular waves which, however, at the point of observation or detection have wavefronts that are effectively planar over a given lateral extent of interest.

Historically, to quantitatively characterize elementary phenomena involving interference and patterns of averaged diffracted intensities that arise in the case where illumination is by a fluctuating source (e.g., a thermal as opposed to a laser source), a second-order coherence theory, describing the cross-correlation of the field at two space-time points, was introduced by Mandel and Wolf and others (e.g., [6]).  Both classical and quantum versions of second-order coherence theory have been developed [6].  It is a powerful formalism particularly well-suited for characterizing diffraction patterns resulting from the use of partially coherent light emitted by temporally and spatially extended incoherent sources.  More recently, this theoretical approach has been applied to x-rays [11] and neutrons [8,10] as employed in studies of condensed matter structure on nanometer and Angstrom length scales.  Nonetheless, the meaning of coherence within this theoretical framework, in particular in regard to the cross-correlation function or so-called mutual coherence function, is limited insofar as one of the measures of coherence that we have been discussing -- namely, the transverse width of a wave packet -- is concerned.  This is discussed below.

Be forewarned at the outset, however, that some of the terminology that has become associated with the concept of partial coherence can be potentially misleading.  The term "mutual coherence" function in particular is so fraught.  In the following, it is argued that the mutual coherence function serves primarily as a measure of the degree to which the spatial extension of an i*ncoherent* radiation source diminishes the ability to resolve certain features of a diffracted *intensity* pattern -- a pattern that is itself the result of a summation of distinct intensity contributions, each created in the first place by a *coherent* process involving the superposition of component *amplitudes* originating from a given object illuminated by a single point source (at a sufficient distance that the wavefronts have become effectively planar).  The formalism can be relatively complicated to parse in that attention to subtle distinctions between quantities representing a superposition of amplitudes in contrast to those which are a summation of intensities is required.

The basic meaning of the mutual coherence function and related quantities is typically illustrated through the simple example of how the interference pattern produced by a pair of apertures -- illuminated by perfectly monochromatic and spatially coherent light (Young's experiment) -- differs from that which arises if the waves are quasi-monochromatic and emanate from an extended incoherent source as depicted in Figure 8 (the modulating envelope due to diffraction by each of the individual slits is not shown in the intensity plot of the figure).  This classic experiment is described in numerous texts on optics, but we follow more closely the descriptions given in Born and Wolf [5], Hecht [7], and Mandel and Wolf [6].

To begin, several simplifying assumptions and approximations are made that are valid for the specific case of interest at hand, namely, the elastic scattering of neutrons that originated in a temporally and spatially extended incoherent source such as a liquid hydrogen moderator at a continuous flux reactor.  We will also ignore neutron polarization and consider only scalar wave functions as opposed to spinors (analogous to neglecting the polarization of electromagnetic radiation).  The usable source intensity can be considered, for the sake of argument, sufficiently weak that only one neutron at a time interacts with the diffracting object and the measuring instrument (diffractometer or reflectometer).  In typical circumstances, a single neutron may pass from source to diffracting object to detector before another neutron arrives.  This means that any interference phenomena resulting from the interaction of a single neutron and a diffracting object arises strictly in terms of the component basis states of which the wave packet representing that particular neutron is composed.  It is also a well-established quantum phenomenon that the observation of the diffraction pattern is stochastic in nature, requiring a sufficient number of apparently random detections of individual neutrons across a range of angular or spatial positions to reveal the underlying form of the pattern.

Further, we assume that the neutrons are quasi-monochromatic. This means that each corresponding wave packet possess a temporal coherence length (which is often referred to as a longitudinal spatial coherence length along the mean direction of propagation) that is significantly larger than its transverse coherence length (perpendicular to the mean direction of propagation) -- similar in form to that which we adopted in preceding sections. Here we are primarily concerned with the transverse coherence length or, more accurately, the transverse coherent extent of the wave fronts within the neutron wave packet over which the phase is constant and can thereby give rise to coherent scattering from an effective average material density. We also assume stationarity and ergodicity, i.e., that any time average over a collection of similar neutron quantum particles successively interacting with the scattering object is time independent and that the time average is essentially equivalent to the ensemble average, respectively.

With the simplifications and assumptions so stated, consider again the diffraction arrangement schematically represented (in two dimensions) in Figure 8. Let the distances between source and opaque barrier (including apertures) and detector plane be sufficiently large and the source width small enough that the far-field or Fraunhofer limit is a valid approximation for describing the diffraction. Circular wave fronts emanate from the source at a point $S_0$ along the axis of symmetry, equidistant from each aperture, and travel towards the opaque barrier, becoming nearly planar along the way.

At the barrier, the two apertures act as coherent secondary sources which radiate circular waves which subsequently interfere at the detector plane to produce a well-known interference pattern of detected intensity -- once an adequate number of single neutrons are emitted and diffract in similar fashion (as indicated previously, the modulating envelope due to diffraction by each of the individual slits is not included in the intensity plot of the figure). If all the neutrons were emitted from the same central source point, the diffraction pattern would be perfectly resolved and the so-called "fringe" visibility $V$ defined in terms of the normalized contrast (difference) between intensity maximum $I_{MAX}$ and minimum $I_{MIN}$ (i.e., the minimum immediately adjacent to the maximum) would be greatest and equal to unity:

(5)                     $V = (I_{MAX} - I_{MIN}) / (I_{MAX} + I_{MIN})$

Note that the central maximum of the pattern coincides with the position of the horizontal symmetry axis.

Consider now a wave train emanating from another source point S, one that is off the symmetry axis and not equidistant from either aperture. Once again, the two apertures act as secondary sources of circular waves that produce a similar diffraction pattern -- but one which is now, on the whole, translationally shifted along the vertical detector axis relative to that associated with the central source point, as shown in Figure 8. The sum of the two intensity contributions are also shown in Figure 8 and clearly indicate a loss of fringe visibility or resolution as defined by Equation 5. In effect, extending the source obscures the interference pattern to some degree relative to what would have been observed with a single point source. This does *not* imply, however, that a wavefront of the radiation emanating from any *one* point source has a diminished transverse extent over which the phase is uniform.

The resultant neutron wave function $\Psi_D$ (here the neutron wave function replaces the scalar electric field in an analogous description for visible light) at a point D on the detector plane is the sum of the two waves that emanated from each of the two secondary source points or apertures $A_1$ and $A_2$ at that point D and is given by

(6) $$\Psi_D (t) = C_{A1} \Psi_{A1} (t - r_1/v) + C_{A2} \Psi_{A2} (t - r_2/v)$$

where t is a point in time, v the phase velocity of the wave, and the (time-independent) coefficients $C_{A1}$ and $C_{A2}$ account for changes in the wave functions that depend upon which aperture the respective component emanated from. In other words, Equation 6 says what the wave function or field amplitude is at any given position and time on the detector in terms of what the wave functions at apertures $A_1$ and $A_2$ were at earlier times $t_1 = r_1/v$ and $t_2 = r_2/v$, respectively.

Ultimately, what is observed on the detector is an intensity pattern acquired over a finite time period $\tau$ which is taken to be long in comparison to the coherence time (which is simply related to the longitudinal or temporal coherence length that was discussed earlier). The net intensity $^\Sigma I_D$ is obtained by averaging over the finite time interval T and accounts for neutrons emanating from every possible point on the extended source. This average is denoted by:

(7) $$^\Sigma I_D = < \Psi_D (t) \Psi_D^* (t) >_T$$

Substituting the explicit expression for $\Psi_D$ given in Equation 6 into Equation 7, expanding, changing variables to $\tau = t_2 - t_1$ (imposing the condition of stationarity), and identifying quantities $I_{D1}$ and $I_{D2}$ as corresponding to the intensities which would be obtained at point D if only either aperture A1 or A2 alone had been open, respectively, we obtain (see, for example, Section 12.3, [7])

(8) $$^\Sigma I_D = I_{D1} + I_{D2} + 2 (I_{D1} I_{D2})^{1/2} Re \{\Gamma_{A1A2} (\tau) / [(\Gamma_{A1A1} (0)\Gamma_{A2A2} (0)]^{1/2}\}$$

$$= I_{D1} + I_{D2} + 2 (I_{D1} I_{D2})^{1/2} Re [\gamma_{A1A2} (\tau)]$$

where $\Gamma_{A1A2} (\tau) = < \Psi_{A1} (t + \tau) \Psi_{A2}^* (t) >_T$, $\Gamma_{A1A1} (0) = <|\Psi_{A1} (0)|^2>_T$, and $\Gamma_{A2A2} (0) = <|\Psi_{A2} (0)|^2>_T$.

The quantity $\Gamma_{A1A2}(\tau)$ in the interference term which arises upon superpositon of the waves is a two space-time point cross-correlation function that is commonly referred to as the mutual coherence function [6]. The argument of the Re part of the function in the last term on the RHS of Equation 8, $\gamma_{A1A2} (\tau)$, is called the normalized mutual coherence function or the complex degree of coherence -- its modulus can be shown [6] to be identical to the fringe visibility given by Equation 5 (see also, [7], Section 12.3) which varies from zero (for complete loss of contrast) to unity (for the optimum contrast possible as would be obtained with a perfectly spatially coherent monochromatic [single point] source). For values in between zero and one, the diffraction pattern manifests some degree of partial fringe visibility.

Up to this point in the discussion, no explicit form for the wave function $\Psi(r,t)$ in the formula for the mutual coherence function has been assumed. Although localized wave packet functions are ultimately of interest, it happens that a circular (in 2D) or spherical (in 3D) wave, which at a sufficient distance from its point source is effectively planar, suffices for describing the fundamental meaning of the mutual coherence function. In analogy to classical light optics, we take $\Psi(r,t)$ to play the role of the electric field intensity for unpolarized electromagnetic radiation. As it turns out, to relate the fringe visibility or instrumental resolution to the relevant characteristics of the radiation source, it is necessary only to evaluate the informational content of the normalized mutual coherence function or complex degree of coherence $\gamma_{A1A2}(\tau)$. The mathematical details are given in Appendix A with the result that the complex degree of coherence is given in terms of the source size 2s and aperture spacing 2a (see Figure C1 in the appendix for a pictorial labeling of these quantities) by

(9)  $\gamma_{A1A2} = \sin\left[(2s)(2a)\pi/(l\lambda)\right] / \left[(2s)(2a)\pi/(l\lambda)\right] = \mathrm{sinc}\left[(2s)(2a)\pi/(l\lambda)\right]$

which can be explicitly related to the fringe visibility V defined in Equation 5 (for the case of the two-slit interference pattern) (see, for example, [7], Chapter 12) --

(10)  $V = (I_{MAX} - I_{MIN}) / (I_{MAX} + I_{MIN}) = |\gamma_{A1A2}| = |\mathrm{sinc}\left[(2s)(2a)\pi/(l\lambda)\right]|$

where $0 < |\gamma_{A1A2}| < 1$ (a well-known result in light optics).

So what exactly does this expression for the complex degree of coherence or fringe visibility tell us? The modulus of $\gamma_{A1A2}$ given by Equation 10 is plotted in Figure 9. From Equation 10, the resolution of the features in the interference pattern depends upon source size (2s), aperture spacing (2a), the distance between source line and aperture line, l, and the wavelength, $\lambda$. For a given wavelength and aperture spacing and a fixed distance between source and apertures, the fringe visibility decreases with increasing source size -- or, equivalently, with an increasing range of the angular distribution of trajectories of wavevectors directed from the source points towards the apertures. That is, the greater the angular divergence of the incident *beam*, the poorer the resolution of the features of the interference pattern. It is straightforward to explicitly relate the argument of the complex degree of coherence or the mutual coherence function to the beam angular resolution. If we take the value $\pi$ of the argument of the sinc function of Figure 9 where the first zero occurs (and beyond which the features of the interference pattern are significantly and progressively further diminished), then we can write

(11)  $(2s)(2a)\pi/(l\lambda) = \pi \Rightarrow$

  $[2\pi / (k\,\Delta\theta_{SOURCE})] \approx (2a)$

where $\Delta\theta_{SOURCE} \approx$ (s/l) and k = $2\pi/\lambda$. The quantity k $\Delta\theta_{SOURCE}$ is a measure of the width of the distribution describing an uncertainty in neutron wavevector components transverse or normal to the mean direction of propagation (neglecting the relatively insignificant contribution from the uncertainty in the magnitudes of the wavevectors, i.e., the parallel or longitudinal parts, for the present case where the radiation is nearly monochromatic -- i.e., $\Delta k/k$ is typically 0.01 in pertinent scattering measurements). This implies that to resolve a spatial dimension of the order of 2a (the separation of the two points A1 and A2), the instrumental beam resolution must be of the order of k $\Delta\theta_{SOURCE}$ = [$(2\pi)$ / (2a)] -- which is the conventional picture. We can, therefore, rewrite Equation 9 in the form

(12)  $\gamma_{A1A2}$ = sinc [(2s)(2a)$\pi$/(l$\lambda$)] = sinc [$2\pi \Delta\theta_{SOURCE}$ (2a)/ $\lambda$]

or  | $\gamma_{A1A2}$ | = |sinc [$2\pi \Delta\theta_{SOURCE}$ (2a)/ $\lambda$]| = V = ($I_{MAX}$ - $I_{MIN}$) / ($I_{MAX}$ + $I_{MIN}$)

This relationship between fringe visibility or the mutual coherence function and geometrical angular beam resolution is widely accepted and not contended here. Note, however, that this does *not* imply that the nearly planar wave train emanating from any *one* source point by itself creates a less-resolved diffracted *intensity* pattern -- it is simply hidden among the similar but shifted patterns contributed by all the other source points which are emitting at the same time. It is this summation of the intensity contributions from a collection of such source points on the extended source line that causes the resolution of the net resulting pattern to be effectively diminished. In this sense, the mutual "coherence" function -- as shown to be associated with the fringe visibility -- can be thought of more as a measure of the degree to which increasing the spatial extent of the source effectively obscures the underlying pattern that would have been better resolved by restricting the source size to one point. Once again, this calculation of the mutual coherence function was performed, as typically done, assuming an explicit plane wave form for the wave function where the spatial extent of a wavefront of constant phase is infinite. Nonetheless, the resultant expression is essentially unchanged regardless of the specific form of the wavefunction for the quantum particle emanating from the source, so long as it is sufficiently planar upon arrival at the object.

Thus, the visibility of a diffraction pattern can be predicted in terms of what would be observed for scattered radiation emanating from a hypothetical pair of points on an object a distance (2a) apart when illuminated by incident radiation from all of the points on a primary spatially extended incoherent source. For any specified degree of visibility V between zero (completely unresolved) and one (maximum resolution), the mutual coherence function predicts what the largest lateral dimension (2s) of the primary incoherent *source* can be. Again, implicit in this determination is the assumption that every wavefront emitted from any single point on the extended source has a phase uniform to within one wavelength at the location of the two points a distance (2a) apart on the diffracting sample object.

As a specific example, the solution of the transcendental equation represented by Equation 12, namely,

(13)  V [(2s)(2a)$\pi$/(l$\lambda$)]= |sin [(2s)(2a)$\pi$/(l$\lambda$)]|

gives, for a fringe visibility V of 0.1, λ = 5 Å, l = 2 m, and (2a) = 2.25 μm, a maximum source dimension of approximately 200 μm.  But, at the same time, it is implicitly required that the wavefront of any neutron emanating from the source and incident on the sample object have a uniform phase (again, to within a wavelength) across a transverse distance of at least 2.25 μm.

In summary, the mutual coherence function primarily describes the degree to which the superposition of intensity contributions from different source points diminishes the ability to resolve features in the intrinsic diffraction pattern for a given object (which would have been completely resolved if illuminated by a single ideal point source).  It is a useful function in that it quantifies the effect of beam angular divergence (which depends upon the distance l and source size 2s) on the instrumental resolution.  On the other hand, the mutual coherence function is not a direct measure of the transverse uniformity of a wavefront in a packet associated with any one individual quantum particle in a beam originating from any particular source point.  And it is the transverse uniformity of the wavefront which determines whether a pair of apertures, for example, can be illuminated in phase to begin with.

Coherent Averaging by a Wave Packet

There is another type of measurement that can be more sensitive to the extent over which the phase of a wavefront is uniform.  Through specular reflection at glancing angles of incidence, the structure and composition of a material object such as a patterned thin film diffraction grating on a flat surface can be tailored to serve as a probe of an individual wave packet's physical shape or spatial extent.

It is well known, from scattering theory as well as in practice, that for specular reflection (wavevector transfer parallel to the mean surface normal) from the surface of a material object wherein the distribution of scattering length density (SLD) varies in a plane parallel to the surface, the effective scattering density is the in-plane average.  In the Born approximation for specular scattering, the reflection amplitude $r_{BA}$ can be written as

(14)                    $r_{BA} = 4\pi/(iQ) \int_{-\infty}^{+\infty} <\rho(x,y,z)>_{xy} \exp(iQz)\, dz$

where ρ is the scattering length density (SLD) of a material sample object, $\mathbf{Q} = \mathbf{k}_F - \mathbf{k}_I$ is the wavevector transfer which is parallel to the z-axis along the sample normal (Q is implicitly taken to be the value along the z-axis in this case), and

(15)                    $<\rho(x,y,z)>_{xy} = (1/A) \int_{-X/2}^{+X/2}\int_{-Y/2}^{+Y/2} \rho(x,y,z)\, dx\, dy = \rho(z)$

where A = XY is the area over which an incident neutron wavefront simultaneously interacts at constant phase.  It can be rigorously shown that this coherent averaging applies to scattering described by exact solutions of the Schroedinger wave equation as well.  (As a related example, it is also straightforward to show that for a given reflection from a periodic crystal lattice, each of the atomic planes contributing to that reflection -- all of which are perpendicular to the corresponding reciprocal lattice vector and wavevector transfer -- has an effective scattering density proportional to the average scattering amplitude of all the atoms lying in that particular plane.)

As already discussed, the spatial extent over which the wavefront is uniform (to a sufficient degree for that required in a given case) is a fundamental quantity in determining the nature and degree of the coherent scattering that is possible from a given object. Figure 10 schematically represents a wave packet similar in form to the elongated wave train described in a previous section but of rectangular cross section (conceptually useful as a simple image of successive wavefronts of constant phase) interacting with a planar sample of inhomogeneous SLD (two values, $\rho_A$ and $\rho_B$). This picture shows that, for elastic, coherent, specular scattering, the distance on the scattering surface that a wavefront of constant phase "sees" along the horizontal axis in the figure, L, is proportional to its transverse dimension $\Delta r_T$ projected a length L across the surface. For the mean wavevector $\mathbf{k}_I$ making a glancing angle $\theta$,

(16)                                $\Delta r_T = L \sin \theta$

The other, orthogonal width (along an axis perpendicular to the plane of the figure itself) in the plane seen by the wavefront is not amplified but equal to whatever the packet width is in that direction.

As given by Equation 15, the effective SLD for the area of the sample seen by the packet with a transverse dimension $\Delta r_T$ is the average density within the projected in-plane area, i.e., a properly weighted area average of $\rho_A$ and $\rho_B$ (pictorially, some combination of blue and red => some shade of purple). Note that this average SLD pertains only to any resultant specular scattering which occurs where the wavevector (and momentum) transfer is strictly perpendicular to the mean surface normal. (Non-specular scattering can also occur at other angles but is not of relevance to the present discussion.) Conversely, if the projected length L were sufficiently less than the dimensions of the areas corresponding to a single scattering length density, either $\rho_A$ or $\rho_B$, then the specular scattering would be observed to be a properly weighted incoherent sum of two independent reflected intensities, each associated with one or the other separate homogeneous region of SLD.

In the next section it is shown how specular mirror reflection at glancing angles of incidence from periodic thin film diffraction gratings of known structure can be used to probe and deduce the transverse extent of a neutron wave packet.

Reflection from Grating Structures at Glancing Angles

Referring again to Figure 10, imagine that the materials of two different SLDs, $\rho_A$ and $\rho_B$, are rearranged to be of uniform (and equal) width and spacing along the horizontal x-axis to form an alternating periodic grating structure (with continuous bars of material along the y-axis perpendicular to the plane of the figure). For elastic specular reflection from such a periodic grating at glancing angles of incidence (typically a fraction of a degree for neutrons of 5 Angstrom wavelength), the wavevector transfer $Q = 2k_M \sin\theta_M = Q_z$ is along the z-axis, perpendicular to the (x,y)-plane of the grating surface. The mean wavevector of a neutron packet, $\mathbf{k}_M$, has an incident (I) and final (F) direction, prior to and after scattering, respectively, whereas the magnitudes of both wavevectors are equal since the scattering is elastic.

The ensemble of similarly shaped packets composing the beam have different mean wavevector directions relative to an average value as characterized by the angular range $\Delta\theta_{BM}$ (which can be defined as the FWHM of such a distribution). As originally depicted in Figure 1, this instrumental beam angular divergence $\Delta\theta_{BM}$ is distinct from the $\Delta\theta_{WP}$ which corresponds to the intrinsic transverse wavevector uncertainty $\Delta k_{WPT}$ associated with an individual packet. The instrumental beam angular divergence corresponds to a directional distribution of N mean packet wavevectors $\mathbf{k}_M$, each of which has a one-to-one correspondence with a specific one of the N individual neutrons and their associated state or packet wave functions within the ensemble composing the beam. Although both angular distibutions, $\Delta\theta_{BM}$ and $\Delta\theta_{WP}$, can contribute to the instrumental angular resolution, only $\Delta\theta_{WP}$ is directly connected with the transverse coherent extent of a neutron packet wavefront $\Delta r_T$ through the uncertainty relation $\Delta k_{WPT} \, \Delta r_T \geq 1/2$. As will be discussed below, gratings of known periodicity and structure can be used to infer $\Delta r_T$ for neutron wave packets ([4] and references therein) and have also been employed in studies of x-ray coherence [22,23,24].

In terms of wavevector transfer Q, the conventional measure of the instrumental resolution for a typical neutron reflectometer along the z-axis normal to the plane of the grating structure, $\Delta Q_z$, is given by

(18) $\qquad\qquad\qquad \Delta Q_z / Q_z \approx [(\Delta\lambda_{BM} / \lambda_{BM})^2 + (\Delta\theta_{BM} / \theta_{BM})^2]^{1/2}$

where $\Delta\lambda_{BM} / \lambda_{BM} = \Delta k_{BM} / k_{BM}$, the subscript "BM" indicating a beam mean or average of the individual $\lambda_M$ or $k_M$ mean packet values. The first term on the RHS of the above equation represents the spread in wavelength or wavevector magnitude in the beam ($\Delta\lambda_{BM} / \lambda_{BM} \approx 0.01$ for a typical reflectometer) whereas the second term describes the degree of geometrical angular divergence in the beam. We are primarily concerned with this latter term as discussed in the preceding sections. For a given magnitude of $k_M$, the range of $Q_z$ due to the beam angular divergence, is, in the small angle approximation, given by

(19) $\qquad\qquad\qquad \Delta Q_z \approx 2 k_{BM} \, \Delta\theta_{BM}$

where $\Delta\theta_{BM}$ is the geometrical angular divergence of the monochromated beam defined by a pair of slit apertures of appropriate width and separation distance such that $\Delta\theta_{BM}$ is of the order of a few minutes or seconds of arc and for glancing angles of incidence $\theta$ of the order of a few degrees at most. (In such a low-angle limiting case, the distribution of transverse components of the packet mean wavevectors within the beam relative to the average mean wavevector direction is nearly the same relative to the reflecting surface.) In typical practice, the instrumental resolution at the critical edge for total external or mirror reflection as well as at the positions of the first few Kiessig fringes (due to the finite thickness of the film bars along the surface normal) observed for specular reflection from a thin film of finite thickness is well-approximated by Equation 18.

It is also possible, in principle, to extract the broadening of the intrinsic grating line width due to the finite transverse size of a packet from the observation of non-specular scattering along the x-direction as a function of $Q_X$ (corresponding to diffraction from the periodic structure of the grating lines) -- assuming the geometrical beam angular divergence is known and the periodicity of the grating is of sufficiently long range order. However, as will be discussed in the following section, obtaining substrates of the requisite flatness is in practice problematic [4]. As a consequence, only specular reflection measurements will be considered here.

Figure 11 summarizes how the position of a critical edge for total external mirror reflection can be used as an indicator of the projected coherent extent of neutron packet wavefronts. If the projected wavefront is of sufficient extent to effectively average over the two SLD values, one associated with the grating bars and the other with the troughs, then the specular reflection corresponds to a coherent scattering process for a material with a uniform SLD that is the average of that of the bar and trough. If, on the other hand, both the widths of the bar and trough are each sufficiently larger than the neutron wavefront's projected dimension, than the observed specular reflectivity will represent the area-weighted incoherent sum of the reflected intensities for the bar and trough separately. Also shown in Figure 11 are the principal experimental results summarizing earlier work [4]. Relevant critical Q values are given in Table 3.

To better illustrate the different roles that the conventional instrumental beam resolution and the finite transverse extent of an individual neutron packet wavefront have on the observed specular reflectivity, model calculations were performed for different instrumental beam resolutions in the two limiting cases: 1) the transverse dimension of the wavefront is of sufficient extent to completely average over a large enough number of the bars and troughs of the grating structure; and 2) the widths of the bar and trough are each significantly larger than the projected transverse extent of the neutron wavefront (bar and trough widths are equal). The substrate was taken to be silicon with approximately 950 Angstrom thick nickel bars deposited on top. Neutrons were incident from vacuum. Figure 12 shows plots for both the coherent average and incoherent sum cases at two extremes of instrumental angular beam divergence, approximately $3.5 \times 10^{-5}$ and $1.5 \times 10^{-3}$ radians (and for a fractional wavelength resolution of 0.01).

Note that in Figure 12, despite a difference of a factor of over 40 in the angular divergence of the beam and the consequential rounding of the critical edge and smearing of the film thickness or Kiessig intereference fringes at the broader angular divergences, an unambiguous distinction between the cases for coherent averaging and incoherent summation still can be made. This clearly demonstrates that the beam angular divergence and associated instrumental resolution along the z-axis normal to the film surface can be measured separately from the transverse extent of the wavefronts within an individual neutron wave packet. Figure 13 shows in more detail the reduction in the Kiessig fringe visibility with broadening beam angular divergence.

It was found in earlier work [4] that by measuring the specular reflectivity from a set of gratings over a range of different periods, the transverse dimension of a packet wavefront is of the order of a micron. In the present work reported here, more precise measurements of the specular reflectivity were performed on the 10 micron (950 Angstrom thick Ni bar width) + 10 micron (trough width) = 20 micron (period) grating as a function of beam angular divergence. The standard neutron reflectometer configuration was employed, the essential components of which are depicted in Figure 1, but where in between the HOPG(002) ideally imperfect mosaic crystal monochromator and neighboring slit was inserted a polycrystalline Be filter (to remove higher-order neutron wavelengths). A guide tube emanating from the liquid hydrogen moderator cold source at the NCNR illuminated the PG. In addition, however, the exit slit prior to the grating sample (which would be located just to the right in Figure 1) was positioned in close proximity to the sample to ensure that the beam footprint was fully intercepted by the 7.5 cm diameter Si substrate upon which the thin film grating structure was deposited, even at the largest beam divergences.   A detector was positioned at a scattering angle (equal to twice the glancing angle of incidence) downstream.  Figure 14 is a composite plot of the measured specular reflectivites as a function of beam angular divergence (in all cases $\Delta\lambda_{BM} / \lambda_{BM} \approx 0.01$) for the 10 + 10 = 20 micron grating with the mean beam wavevector perpendicular to the grating bars.  At all angular beam divergences, including the widest, the individual neutron packets effectively average over bars and troughs resulting in a single critical cutoff corresponding to the average SLD of the Ni bars and the empty (except for air) spaces in between.  On the other hand, it is found that even for the narrowest angular beam divergence the bars and troughs of a grating with a 20 + 20 = 40 micron period (only twice as long) are *not* averaged over by the neutron.

Previous measurements [4] performed with the stripes *parallel* to the beam direction  -- where no amplification of length by projection at glancing angles occurs -- yielded an incoherent sum of separate reflectivities even for the smallest grating repeat period available at the time (5 + 5 = 10 microns).  Nonetheless, the case for stripes parallel to the nominal mean wavevector is potentially more complicated below the critical Q (for total external mirror reflection) where the Born approximation is no longer valid -- what gets averaged over may depend on different penetration depths for the neutron wave function incident on a stripe as opposed to a trough.

Table 4 lists typical reflectometer slit widths and corresponding beam angular divergences corresponding to the data shown in Figure 14 along with the geometrical angular resolution for the incident beam.  The geometrical angular widths (FWHM) calculated from the slit widths and their separation distance are typically found to be consistent with measured values to within a few (2 to 3) percent accuracy for slit widths approximately 0.1 mm or greater.

Note that if the coarsest instrumental (geometrical) beam resolution listed in Table 4 had been used in the uncertainty relation (Equation 2)  -- as previously shown to be inappropriate -- it would predict the transverse coherent dimension of an incident neutron wavefront $\Delta r_T$ to be $1 / 2\Delta k_{BMT} = 0.0269$ µm.  This value would be far too small to average over the stripes and troughs of the 20 µm period grating and would be in contradiction to that indicated by the data shown in Figure 14.  Once again, evidence shows that the packet $\Delta r_T$ is *not* obtained from the distribution of transverse components of mean packet wavevectors which defines the geometrical angular divergence of the incident beam (where $\Delta k_{BMT} \approx k_{BM} \Delta\theta_{BM}$ and $\Delta\theta_{BM} \approx \arctan[(W_1 + W_2) / (2L_{12})]$).

Figure 15 shows model specular neutron reflectivity curves about the effective critical angle for a grating with neutron wavevector perpendicular to the stripes at two extremes of instrumental beam resolution and where the glancing angular dependence of the projection of $\Delta r_T$ , given by Equation 16, was explicitly taken into account.  The calculation used to generate the model specular NR curves in this figure yields exact solutions of the one-dimensional, time-independent Schroedinger equation -- but employing plane wave functions.  The range over which averaging performed by a packet wavefront of finite extent is imposed independently.

Like the Wavy Surface of a Circus Mirror

In preceding sections, measurements of the transverse width of the neutron packet wavefront via diffraction from phase gratings at normal incidence as well as by specular reflection from thin film gratings at glancing angles were described in which markedly different results were obtained. Using phase gratings, in the case of the work reported herein as well as that of others [19], the uniform transverse extent of the packet wavefront was found to be of the order of tens of microns in contrast to a value of the order of a micron or less in the method involving specular reflection. Why the apparent disparity? After all, the instruments in all cases employed an HOPG(002) pre-monochromator at nearly the same neutron wavelength and with comparable geometrical angular divergences of several seconds of arc, as described earlier. It is true that the USANS instrument used by Treimer et al. [19] also employed multiple-bounce, channel-cut Si(111) monochromator and analyser, but as the calculations as well as measurements presented here indicate, the HOPG(002) reflection alone probably suffices to impart to the neutron packet a $\Delta r_T$ of the order of tens of microns.

The reason for the discrepancy may have to do with the flatness of the supporting Si substrates on which the Ni-stripe reflection gratings are deposited. In the case of specular reflection from thin film patterned structures at low wavevector transfer (glancing angles of incidence relative to the reflecting surface), curvature of the underlying support substrate can effect the measurement of the incident neutron packet $\Delta r_T$ and distort that of the specularly reflected one, as shown in the following illustrative example.

Figure 16 shows a simple representation of a wavy surface as a sinusoid. For a plane wavefront of width W incident from the left, the lower edge making contact with the surface first will be out of phase by π -- relative to the specularly reflected wavefront -- when the path length difference is half a wavelength. The surface height variation $\Delta_\pi$ can be determined to be

$$(20) \qquad\qquad \Delta_\pi = \lambda\, S\, / \,(4\, W)$$

where S is the distance across the surface as depicted in the figure and λ is the neutron wavelength. The maximum angle of tilt α for a local surface normal from the mean surface normal is approximately given by

$$(21) \qquad\qquad \tan(\alpha) = \Delta_\pi\, / \,(S/2) = \lambda\, / \,(2\, W)$$

(α could also be taken to represent the width of an angular distribution). The glancing angle of incidence θ is given by

$$(22) \qquad\qquad W = S \sin(\theta)$$

so that for W =0.5 microns and θ = 0.5 degrees (critical angle for Ni at λ = 5 Å) S = 57.3 microns (similar to the values associated with the measurements described earlier). For λ = 5 Å and W =0.5 micron, α is approximately 5. x $10^{-4}$ radians or 0.029 degrees -- which roughly corresponds to what would be found for a polished silicon substrate typically employed in neutron reflectometry (and also for neutron guide surfaces). This particular example represents but one of numerous possible surface topologies and is presented here only to illustrate qualitatively the potential effect of a non-flat surface in distorting an incident packet wavefront in the specular reflection process.

On the other hand, for a phase grating etched onto a silicon substrate and oriented perpendicular to the incident beam in the transmission geometry, the geometrical or angular amplification by the sine function is suppressed and the surface flatness requirement to maintain uniform phase across an incident wavefront is thereby relaxed.

As a practical consequence, for specular neutron reflectometry studies of layered thin film structures, the supporting substrate can become, operationally, an integral component of the instrumental optics in that its curvature can in effect restrict the surface area over which an incident packet can engage in a coherent specular scattering process.  Moreover, the effect of surface curvature in distorting the wavefronts of a reflected packet can also have consequences -- such as for reflection within neutron guide tubes (which ideally would not have surface distortions as extreme as wavy circus mirrors).

Description of a Neutron Wave Packet According to Standard Quantum Theory

Historically, the so-named "standard" quantum theory (SQM) which has emerged over more than a century of development -- and which is represented in the majority of textbooks on the subject -- has proven to be remarkably successful in predicting physical phenomena on the atomic scale, particularly in describing the physical and chemical behavior of condensed matter.  As is well known, however, the current theory is incapable of explaining several basic phenomena such as the collapse of the wave function.  Moreover, certain interpretations of the theory are still being debated as to their validity.  One particular point of contention concerns the description of the wave packet representing a freely propagating Fermion, such as the neutron, which is of central importance here.  Specifically, the issue has to do with whether the wave packet function of Equation 1 is associated with a single, independent neutron or, alternatively, represents an ensemble of similar neutrons composing a beam in which each of the component plane wave momentum eigenstates corresponds to a neutron with a particular associated wavevector (and energy value).  This question has been considered, for instance, in the textbook by Ballentine ([25], Sec. 9.4, pp. 238-241) and in similar discussions elsewhere [26, 27].  We will briefly outline the essential arguments made in Ballentine [25].

*Pure States, Mixed States, and Density Operators*

As a starting point, we consider Ballentine's general view ([25], Sec. 9.3) regarding a state vector, namely that there are two principal classes of interpretation.  One is that what is believed to be a "pure" state ". . . provides a complete and exhaustive  description of an *individual* system."  (Where the pertinent individual system here is taken to be a freely propagating neutron.)   In the second class of interpretation, a " . . . pure state describes the statistical properties of an *ensemble* of similarly prepared systems." According to Ballentine at least, neither of these two different viewpoints has been universally accepted to be the correct one.  Nonetheless, Ballentine adopts the latter interpretation throughout his quantum mechanics text and offers a theoretical argument as one of his reasons for doing so.  We believe that this argument is fundamentally flawed but reproduce the essence of it below so that we can subsequently show specifically where the disagreement lies.

Ballentine [25] considers, as a specific example, the diffraction of electrons (which, also being Fermions, means that this example would be in essence identical for neutrons as well).  It is presumed that a beam of electrons having some spread in energy is produced by and emanates from an appropriate source for the purpose of scattering from a sample object to create, for instance, a diffraction pattern.  It is postulated that the energy spread of the beam can be accounted for by only one of two assumptions:

"(a) Each electron is emitted in an energy eigenstate (plane wave), but the particular energy varies from one electron to the next;

(b) Each electron is emitted as a wave packet which has an energy spread equal to the energy spread of the beam."

Ballentine then states that "One might expect that these two assumptions would lead to quantitatively different predictions about the interference pattern, and so they could be experimentally distinguished . . . " -- and subsequently purports to show mathematically that, on the contrary, it would *not* be possible to do so.  We first outline his argument and then show why it is an inappropriate description of the system at hand -- both for theoretical reasons and because it is in contradiction with actual observations.

The propagation of the free electrons in the beam is described by Ballentine [25] in one dimension only.  Addressing assumption (a) first, each electron in the beam is taken to have a plane wave function $\psi_k(x,t) = \exp[i(kx-\omega t)]$ with an energy distribution in the beam characterized by the probability density $W(\omega)$ (where energy = $\hbar\omega$).  The state operator is then defined to be

(23)        $\rho = \int |\psi_k><\psi_k| \ W(\omega) \ d\omega = \rho(x, x') = <x|\rho|x'> = \int\psi_k(x,t) \ \psi_k^*(x,t) \ W(\omega) \ d\omega =$

$\rho(x, x') = \int\exp[ik(x-x')] \ W(\omega) \ d\omega$

where the time dependence has canceled out thereby indicating that this is inherently corresponds to a steady state -- with the claim that all observable quantities, including the diffraction pattern, can be calculated from the state function operator $\rho(x, x')$.

Next, Ballentine addresses case (b) and assumes that an individual electron in the beam is described by a wave packet state $\psi_{t0}(x,t) = \int A(\omega) \exp\{i[kx-\omega(t-t_0)]\} \ d\omega$ where $t_0$ denotes the particular time that the packet was emitted by the source.  The energy distribution of the packet is $W(\omega) = |A(\omega)|^2$.  The state function for the beam is then taken to be obtained by integrating over all emission times $t_0$

(24)        $<x|\rho|x'> = \lim(T\rightarrow\infty) \ (1/T) \int\psi_{t0}(x,t) \ \psi_{t0}^*(x,t) \ dt_0$

to obtain

(25)        $\rho(x, x') = \int\exp[ik(x-x')] \ |A(\omega)|^2 \ d\omega$

which is the same result for case (a) in Equation (23).  Thus, Ballentine concludes, the two interpretations cannot be distinguished from one another.

So what could be wrong with this conclusion?  There is nothing strictly incorrect mathematically about the operations Ballentine performs in arriving at the equivalency of Equations 23 and 25 for the two cases (a) and (b).  The problem is that the resulting mathematical expressions do not correspond to a physically realistic description of a freely propagating Fermi quantum particle in the present circumstance -- for the specific reasons, both observational and theoretical, given below.

First, it is experimentally observed that the diffraction process for a beam of non-interacting neutrons -- independent quantum particles -- occurs one neutron at a time.  This has been established in numerous experiments, and not only for neutrons but other Fermions such as electrons as well.  The point by point accumulation of captured electrons on a screen over time to eventually produce a seemingly continuous pattern of diffracted intensity from a double slit is exemplified by the well-known work of Tonomura et al. [28].  Certainly the wave function associated with a single neutron diffracted by an object, say the double slit aperture system of Young's experiment, and propagating on toward the detector cannot be described by a single plane wave state as assumption (a) would require since the probability of locating that neutron anywhere on one of its plane wave fronts as it intercepts the detector screen would then be uniform.  On the other hand, for example, a wave function which more realistically describes the neutron state for a neutron single particle diffracted through a rectangular aperture (in 2 dimensions) has the form (in the far-field limit as it approaches a detector position) (e.g., [7], p. 443]

$$(26) \qquad \psi_k\,(r,t) = \mathrm{Re}\,\{\,(C/r)\,[\sin(k_\perp\,D/2)/(k_\perp\,D/2)]\,\exp[-i(kr - \omega t)]\}$$

where r is the distance from aperture to point of observation, k is the magnitude of the mean packet wavevector, $k_\perp = k\,\sin(\theta)$ perpendicular component of the wavevector (where k is the magnitude of the mean wavevector of the packet and $\theta$ is the deflection angle as in Figure 2), D is the aperture width, and C is a normalization constant.  The sinc function in the expression for this wave packet function effectively modulates the amplitude probability in a way which localizes it and creates the familiar diffraction pattern that is associated with a single slit of finite width -- once again, after a statistically significant number of similar events have occurred.

In addition to the physical evidence described above for rejecting assumption (a) of Ballentine as a possible interpretation, there is further reason to do so based on fundamental principles of SQM which constrain what form a wave function can be to describe a pure state associated with a single Fermi quantum particle.  This is shown in a mathematically rigorous way in the recent work of Berk [16], the relevant key results of which we will summarize below.

The mathematical framework of (non-relativistic) SQM has evolved into its canonically accepted present form wherein single quantum particles are associated with appropriate state or wave functions, as already mentioned in preceding sections.  More formally, these single-particle quantum states can be " . . . represented by physically acceptable solutions of the [time-dependent] Schroedinger equation [SE], assigned to a Hilbert space of continuous space-time functions, or to an associated vector space projectable onto function space in the manner of the Dirac formalism . . . " [16].  Moreover, Berk [16] shows that the Fermion wave packets applicable to scattering problems must be solutions of the time-dependent SE (and consequently *not* stationary states) and that a localized wave packet function in all 3 spatial dimensions (which may be constructed, for instance, of a linear superposition of pure plane wave states) does in fact represent a pure quantum state of a single particle. It is also emphasized (as in the work of Maudlin [29]) that no physical process is known by which a pure state can be transformed into a mixed state -- which thereby implies that it is impossible to prepare a single particle in a mixed state. On the other hand, *mixed states* can be constructed to describe *beams* of independent particles created by appropriate incoherent sources and such mixed states can be described by combining state-defined projection operators into statistical operators. These statistical mixtures in effect provide a means of accounting for a classical probabilistic behavior in addition to the underlying quantum nature of a pure state system [30,16].

More specifically, Berk [16] demonstrates exactly what the differences are between pure and mixed states in their respective formal representations as statistical operators or so-called density matrices. He shows that the statistical operator for a pure state

(27) $$\rho^{(\chi)}{}_{PURE}\,(t) = |\psi^{(\chi)}(t)\!><\!\psi^{(\chi)}(t)|$$

averaged over time in the manner followed by Ballentine [25] (see Equation 24) gives

(28) $$< \rho^{(\chi)}{}_{PURE}\,(t) >_T = \lim(T \rightarrow \infty)\,(1/T)\,\int_{-T/2}^{T/2} \rho^{(\chi)}{}_{PURE}\,(t)\,dt = \rho^{(E)}{}_{MIXED}$$

where the derived mixed state density matrix operator $\rho^{(E)}{}_{MIXED}$ is associated with probabilities which are *not*, in general, the statistical weights of superposition in the originally constructed pure state. In other words, this shows, in particular, that a relevant quantity such as the density operator describing an incoherent ensemble of plane wave states (incoherent statistical mixture) is indeed different from that representing a coherently modulated collection of plane waves (coherent superposition) forming a single particle wave packet function.

Moreover, Berk [16] also analyzes, from first principles, the Young double-slit experiment for Fermions in occupation number (Fock) space. He demonstrates that it is also possible and rigorously correct to describe a *many-state* problem in occupation number space in a similar way to what is more conventionally done to characterize *many-body* (particle) problems. That is, a wave packet constructed of a linear superposition of a large number of plane waves can indeed represent a pure quantum state of a single particle in the relevant Fock space.

As has been discussed earlier, although plane waves and stationary state solutions of the time-independent Schroedinger equation have proven to be of great utility for analyzing a multitude of elastic diffraction and reflection data in both neutron and x-ray scattering measurements, such use should be understood as an approximation to physical reality -- and although of sufficient accuracy for many purposes, in certain applications this treatment falls short, as we have discussed earlier, for example, in properly accounting for the actual volume of a sample which can contribute to a coherent scattering process. The physical meaning of the single-particle wave packet which has been adopted here throughout and by Berk [16] is evidently now widely, though perhaps not universally, accepted. For example, Merzbacher [31] states that " . . . it is not permissible to consider [A(**k**)] [from Equation 1] as a measure of the relative frequency of finding various values of k in a large *assembly* of particles.

Instead it is an attribute of a single particle."  It follows, of course, that experimental verification of such theoretical, probabilistic predictions requires measurements to be performed on a sufficiently large number of *identical* systems.

Conclusions

There are two distinct distributions of wavevectors associated with a beam of freely propagating neutrons as prepared in a typical scattering instrument.  One consists of the wavevectors of the components of a coherent superposition of basis states that constitute each individual neutron's corresponding wave packet function, whereas the other is made up of an incoherent collection of mean packet wavevectors associated with the ensemble of neutrons composing a beam.  Both the distribution of the mean wavevectors of individual packets in the beam as well as the wavevector components of the superposition of basis functions within an individual packet can contribute to the conventional notion of instrumental resolution.  However, it is the transverse spatial extent of packet wavefronts -- over which the phase is sufficiently uniform -- alone that determines the area over which a coherent scattering process with matter can occur.  This picture is shown to be entirely consistent in principle with the formal tenets of the standard quantum theory traditionally used to describe scattering.

In addition to defining the distribution of mean packet wavevectors of the individual neutrons composing the beam, the very same monochromating and collimating devices, such as crystals and slits, may further shape each individual neutron wave packet in directions both transverse and parallel to its mean wavevector.  Nonetheless, it is demonstrated that the two distinct distributions -- one consisting of the transverse components of the collection of packet mean wavevectors in the beam and the other made up of the transverse components of basis wavevectors intrinsic to each individual wavepacket -- can be distinguished from one another experimentally in practice.

Moreover, it is also shown that the morphology of a scattering object under study, for example, deviations from perfect flatness of a substrate supporting a thin film sample, can also affect the transverse uniformity of a neutron packet wavefront, thereby, in effect, acting as part of the instrument optics.

The identification of any commercial product used in the course of the research reported herein does not in any way imply an endorsement thereof.  We thank D. Hussey for the use of the neutron phase gratings.  We would also like to thank R. Cappelletti for many valuable conversations on coherence.

Appendix A: Gaussian Wave Packets

In practice, for a broad range of applications, the elastic scattering of unpolarized x-rays and neutrons (in the approximately 1 to 10 Angstrom wavelength range) by condensed matter can be described at sufficiently low energies by a common stationary state formalism (involving time-independent equations of motion), the main differences between them being the scattering length densities appropriate to the two kinds of radiation. Largely for this reason much of the treatment of neutron scattering as employed in studies of the microscopic structure of condensed matter is adopted from light scattering employing the fundamental concepts of refraction, reflection, and diffraction. As a specific example, consider the case for reflectometry at low-enough glancing angles of incidence from a flat (possibly multi-layered) film in which the material of the film can be treated as being continuous. Typically, the reflectivity for both x-rays and neutrons at such low wavevector transfers is so strong that an exact solution of the time-independent Schroedinger equation rather than the Born aproximation -- and normally assuming plane wave functions -- is required for sufficient accuracy.  For unpolarized radiation, the stationary state equation for the wave function $\Psi(\mathbf{k},\mathbf{r})$ has the generic form

(A 1)
$$-\nabla^2 \psi(\mathbf{k},\mathbf{r}) + q(\mathbf{r})\psi(\mathbf{k},\mathbf{r}) = k^2 \psi(\mathbf{k},\mathbf{r})$$

The effective interaction q($\mathbf{r}$) in equation (A 1) is defined for either situation in terms of a scattering length density (SLD) or $\rho(r)$ -- i.e., the scattering length per unit volume -- where q($\mathbf{r}$) = 4π $\rho(\mathbf{r})$.

For the particular case of x-rays, $\Psi(\mathbf{k},\mathbf{r}) = E_\perp(\mathbf{k},\mathbf{r})$ represents the electric field amplitude normal to the direction of propagation of the electromagnetic wave, and equation (A 1) then stands for the stationary state Maxwell equation

(A 3)
$$\nabla^2 E_\perp(\mathbf{k},\mathbf{r}) + \mu_0 \varepsilon(\mathbf{k},\mathbf{r}) k^2 E_\perp(\mathbf{k},\mathbf{r}) = 0$$

where $\mu_0\varepsilon(\mathbf{k},\mathbf{r})$ = 1 - q($\mathbf{k},\mathbf{r}$)/$k^2$ is the permeability function at point $\mathbf{r}$.  For neutrons, we may multiply both sides of (A 1) by $\hbar^2/2m$, where m is the neutron mass, to recover the stationary state Schrödinger equation,

$$-\frac{\hbar^2}{2m}\nabla^2 \psi(\mathbf{k},\mathbf{r}) + V(\mathbf{r})\psi(\mathbf{k},\mathbf{r}) = E(k)\psi(\mathbf{k},\mathbf{r})$$
(A 4)

with V(r) = ($2\pi\hbar^2/m$) $\rho(\mathbf{r})$ and E(k) = $\hbar^2 k^2/(2m)$ is the constant total energy of the incident free neutron.  For both types of radiation, the momentum of the incident beam (defined in vacuum) is $\hbar\mathbf{k}$, and $k^2 = k_x^2 + k_y^2 + k_z^2$ with k = 2π/λ.

The essential differences between x-ray and neutron reflection within the stationary state, time-independent formalism outlined above reside in the scattering length densities appropriate to the radiations.  For neutrons, the SLD associated with the nuclear interaction for a specific isotope varies in an apparently non-systematic manner across the periodic table of the elements whereas the effective SLD for x-rays is proportional to the number of electrons surrounding a given nucleus.  For both neutrons and x-rays, the SLD is a complex function, in general, where the imaginary part phenomenologically accounts for absorption.  In most cases, neutrons are weakly absorbed by materials, and the SLD can be taken to good approximation to be a real number.  X-rays, on the other hand, usually are strongly absorbed, and typically the real and imaginary parts of the SLD are comparable.  These differences in SLD do not affect the fundamental methods to computing scattered intensities but they do, of course, affect the results.  In the more general case of light scattering, Maxwell's equations for both electric and magnetic fields need to be solved, requiring the introduction of many-body photon (Boson) states, as in the case of laser light.

As discussed in a preceding section, whereas a single plane wave basis state is a solution of both the time-dependent and time-independent Schroedinger equations of motion, the 3D wave packet of Equation 1 satisfies only the explicitly time-dependent version.  For mathematical convenience, as discussed in detail in [16], the free particle Schrödinger wave packet is almost universally described by the "standard" Gaussian model, as specified (in 1D) by the wave vector distribution that provides the amplitudes for the coherent superposition of plane waves, viz.

$$\wp_{coh}(k \mid \bar{k}, \Delta k, t) = \exp[-\frac{(k - \bar{k})^2}{2(\Delta k)^2}]\exp(i\beta k^2 t/2)$$

(A 5)

with $\beta = \hbar /m$.  For continuous k the needed superposition is defined by a Fourier transform (FT) with respect to k, leading to the wave function

$$\psi(x,t) = c(t)\exp[-\frac{(X - 2\beta\bar{k}T)^2}{2\sigma^2(t)}]\exp(-i\chi(x,t))$$

(A 6a)

$$\sigma(t) = \sqrt{\sigma_0^2 + 4\beta^2 T^2/\sigma_0^2} = \sigma_0\sqrt{1 + (\Delta k/k)^2 (X(T)/\sigma_0)^2},$$

(A 6b)

where

(A 6c)                    $\chi(x,t) = 4[\bar{k} X - (k^2 - X^2 / \sigma_0^4) \beta T]$

and

$$c(t) = \Delta k / \sqrt{1 + 2i\beta(\Delta k)^2 T}$$

(A 6d)

with $\sigma_0 = (\Delta k)^{-1}$, and $X = x - x_0$ and $T = t - t_0 > 0$.  (Typically, in textbook discussions, $x_0 = t_0 = 0$.) One may note that for $k_M > 3\Delta k$, as nearly would be the case for almost plane-wave-like wave packets, the Gaussian form essentially enforces the restriction to k > 0 in the FT producing $\Psi(x, t)$. Also, in (A 6b), X (t) = $v_g T$ is the distance traveled by the wave packet in time T, where $v_g = 2 \beta k$ is the group velocity. Thus in (A 6a) and (A 6b) we see the well-known spreading of the (massive) free wave packet with increasing t (or X(T)), the rate of spreading increasing with smaller initial spatial localization (i.e., with greater $\Delta k$). Non-spreading wave packets -- called solitary waves or solitons -- are characteristic of massless particles, a well-known mathematical result of the linear k-dependence of their kinetic energy, and so are not solutions of the Schrödinger equation. Solitons are possible, however, for massive particles in the relativistic realm as special solutions, for example, of the 3-D Klein Gordon equation. As regards solutions of the Schrödinger equation, we are aware of only one example of a true soliton -- but only in 3 dimensions, as a spherically symmetric wave packet (see [16] for relevant citations).

The second equality in (A 6b) notwithstanding, $\sigma(t)$ is independent of $k = k_M$, the mean or "group" wave vector of the packet, implying that a 2- or 3-dimensional wave packet spreads in time in each of its geometrically defined dimensions, dependent only on the value of $\Delta k'$ for each dimension. Specifically, for a 2 D {x, y} wave packet with longitudinal dimension along x and transverse dimension along y, and with the Gaussian model applying along each axis, we now have (A 6) along x, with $\sigma(t) \rightarrow \sigma_x (t)$, $k \rightarrow k_x$, etc.; while along y, $\sigma_0 (t) \rightarrow \sigma_{0y} (t) = 1 / \Delta k_{y'}$, $k \rightarrow k_y = 0$, etc., and now with

(A 7)
$$\sigma_y(t) = \sqrt{\sigma_{0y}^2 + 4\beta^2 T^2 / \sigma_{0y}^2} = \sigma_{0y}\sqrt{1 + \left(\Delta k_y / k_x\right)^2 \left(X(T)/\sigma_{0y}\right)^2},$$

the last formula expressing the broadening along y in terms of the distance traveled along x. It is interesting to notice that since the rate of broadening along orthogonal dimensions increases with the degree of initial localizations along the axes, a cigar shape (elongated) wave packet eventually evolves into a pancake _shaped (flattened) wave packet as it expands, while an initial "pancake" expands into a "cigar."  Substituting typical values of $\sigma_{0y}$ = 12.76 x $10^4$ Å (FWHM = 30 microns) and k = 1.2566 Å$^{-1}$ , Equation A7 gives $\sigma_y$ to be broader than $\sigma_{0y}$ by only about 0.6 % at a distance of 1 meter from the point of creation of the packet -- which is a relatively small degree of spreading.

Wave packet spreading is a purely mathematical result of the quadratic form of kinetic energy for particles of non zero mass, viz., E / ℏ  = β k2 . It is sometimes rationalized by appeal to the uncertainty principle, which, however, could possibly suggest the time dependent constraint σ(t) ~ 1 / Δk(t) . Such behavior, however, depends on the subtlety of whether we are considering the state Ψ(x, t) or its modulus |Ψ(x, t)|.  For example, while the wave packet in (A 6) spreads, the distribution of component k values in Ψ(x, t), as defined by its inverse FT (FT$^{-1}$), necessarily remains unchanged. On the other hand, the distribution of component k values in |Ψ (x, t)|, possessing unit phase, is the FT$^{-1}$ of a Gaussian, thereby producing a Gaussian distribution of k values having standard deviation Δk(t) = σ(t)$^{-1}$.  However, Ψ(x, t) is a pure state -- i.e., a wave packet solution of the time-dependent Schrödinger equation -- while its modulus |Ψ(X, t)| is not. The physical difference is that, according to the Born rule,  |Ψ (x, t)| via its square, relates directly to position measurements of a neutron in the state Ψ(x, t), which only then brings to bear the relevance of the uncertainty principle. The state function Ψ(x, t), however, as discussed in [Berk, arXiv], does not alone imply a measurement outcome until it is projected onto an eigenstate of the "observable" of interest; and thus its pre-measurement k-content is not subject to bounds implied by possible post-measurement implications of the uncertainty principle.

The similarities between neutron and x-ray scattering extend to crystal diffraction at higher wavevector transfers where the discrete positions of atoms in a material can be determined with sub-Angstrom accuracy.  Here the scattering is normally weak enough that the Born approximation (also employing plane wave solutions) can be applied.  In this formalism the scattering potential for individual atoms (described by a scattering length) and their positions within the material are related to the scattering pattern via a Fourier transform.  In this conventional scattering theory (see, for instance, [32]) it is assumed that each neutron-nuclear interaction gives rise to a spherical scattered wave with a certain amplitude (as characterized by a scattering length).  In the case of a purely elastic and "coherent" interaction,  a collection of spatially ordered nuclei within a material object can be stimulated simultaneously, in unison, across an incident neutron wave front of uniform phase to create an outgoing reflected neutron wavefunction -- which, in the far field, approaches a form with nearly plane wavefronts.  In contrast, if the nuclei of the object interact with the incident neutron entirely via an "incoherent" potential (e.g., through a spin-dependent process in which the relative orientation of the spins of the neutron and nuclei are random), then each excited nucleus will give rise to an independent scattered wave that is spherical but in general not in phase with any of the similarly scattered waves from other nuclei.

Nonetheless, at present, the actual shape, size, and composition of the packet embodied by A(**k**) for a neutron immediately upon emission (presumably formed during the process) from a temporally and spatially extended incoherent source -- or exactly how it subsequently evolves in time and space -- is not definitively known.  A considerable body of work on fundamental concepts in quantum theory exists regarding possible theoretical descriptions of spatially localized wave functions and the wave equations which govern them -- perhaps most notably the work of Ghirardi, Rimini, Weber, and Bassi [33].  Their theory introduces stochastic and nonlinear elements to a new dynamics which addresses the measurement problem and collapse of the wave function which the Schroedinger equation does not -- yet approaches the older description in some limit.  Whether or not this particular theory proves to be completely correct, it is widely accepted that the neutron wave function must be spatially localized to a significant degree, enough that it can be confined within various devices such as beam tubes, guides, and instruments including diffractometers and interferometers (e.g., it would be unreasonable to expect that a neutron which scatters from a specific material sample within a given diffractometer could simultaneously be present in a neighboring instrument to interact with a different specimen -- as might be inferred *if* the neutron wave function were in fact a plane wave with wavefronts of infinite spatial extent).  A general discussion of possible neutron wave packet forms relevant to scattering processes as well as fundamental issues pertaining to neutron scattering theory are also discussed, for example, in the text on neutron optics by Utsuro and Ignatovich ([34], chapters 9 and 10, respectively).

Appendix B: Reflection from Perfect Crystal Mosaic Blocks

A typical monochromator crystal such as highly oriented pyrolytic graphite (HOPG) -- made up of a mosaic or orientational distribution of perfect micro-crystalline blocks -- can, in practice, shape both an individual packet momentum distribution as well as the wavelength spectrum (and to a limited extent the angular distribution) of the beam as a whole.  A typical micro-crystallite, with dimensions of the order of several microns, reflects a single neutron by a coherent elastic process [Bragg diffraction from the (002) atomic planes] that can alter or limit the wavevector components of an individual packet.  On the other hand, the incoherent sum of such reflections of different neutron packets from various mosaic blocks that are at dissimilar angles relative to a mean direction (the mosaic blocks are oriented according to a normal angular distribution that is typically of the order of a half a degree FWHM) define a beam which has a distribution of individual neutrons with their particular packet mean wavevectors.  As we have been discussing, it is this distribution of the mean wave vectors of the individual packets constituting a beam that is often the primary contribution to the more conventional notion of instrumental resolution.

But it is also shown in this appendix that a mosaic crystal block can define the size and shape of a reflected neutron wave packet in two distinct ways.  The finite in-plane dimensions of the stack of reflecting atomic planes as well as the number of contributing reflecting planes can both affect the transverse extent of the wavefronts of the reflected neutron packets.

Before examining the scattering process for a single crystalline mosaic block according to a more traditional treatment of crystal diffraction, it is instructive to perform the same type of Huygens-Fresnel wavelet construction as was applied to the aperture in a preceding section.  Wavelets are taken to emanate from each source point (an atom) isotropically (multiple scattering processes are neglected).  Shown in Figure B 1 is the result of such a calculation assuming a generic crystal 10 microns wide with 100 reflecting atomic planes spaced of 5 Å apart from each other, and with 1000 atomic source points per plane. The wavelength was taken to be 5 Å as well and the distance between crystal face and point of observation of the reflected wave was 0.5 meter. This two-dimensional block of source points was taken to be illuminated by plane waves in phase such that the Bragg diffraction condition was effectively satisfied. As in the case of an aperture of the same width, the reflected wave has a well-defined lateral dimension which at 0.5 m from its source has a uniform wavefront (to within one wavelength) over a lateral extent of approximately 22. microns -- similar to that produced by the single aperture of the same width. Thus, the finite lateral size of the reflecting crystal block can affect the lateral dimension of the reflected neutron wave function.

But, in addition, the number of diffracting crystal planes within the crystal block can also contribute to the shaping of a reflected neutron packet. Consider the diffraction from a single mosaic block in the Born approximation in which multiple scattering is neglected and assume, for the time being, a single plane wave to be incident. The crystal is taken to be that typically used as a neutron monochromator, pyrolytic graphite. The scattering geometry is depicted in Figure B 2. The incident and scattered wavevectors are $\mathbf{k_I}$ and $\mathbf{k_F}$, respectively, while $\mathbf{Q} = \mathbf{k_F} - \mathbf{k_I}$ is the wavevector transfer ($|\mathbf{k_I}| = |\mathbf{k_F}| = k$ since the scattering is elastic). For the specular condition of interest here, where glancing angles of incidence and reflection are of equal magnitude, the momentum transfer is normal to the reflecting atomic planes and no momentum transfer occurs along the x- and y- in-plane directions -- thus $Q = Q_Z$.

The coherent reflectivity $|r_{BA}(Q_Z)|^2$ for a set of N parallel atomic (002) planes in a perfect single crystalline mosaic block of hexagonal pyrolytic graphite is, in the Born approximation, given by

(B 1)                         $|r_{BA}(Q_Z)|^2 = (4\pi / Q_Z)^2 \, \rho^2_{PG(002)} \, [\sin(NQ_Zd/2) / \sin(Q_Zd/2)]^2$

where $\rho_{PG(002)}$ is the scattering density associated with the (002) atomic planes of spacing d. The FWHM of the central peak that is commonly associated with the (002) Bragg reflection is approximately

(B 2)                         $\Delta Q_Z = (2\pi) / (Nd)$

where the reflected intensity maximum occurs at

(B 3)                         $Q_{Z\,MAX} = 2\pi / d$

In general, $Q_Z = 2k\sin\theta_I$ where $\theta_I$ is the angle of incidence relative to the (002) atomic planes. Equation B 2 indicates an inverse relationship between the range of allowable incident angles over which reflection can occur and the number of atomic planes contributing to the reflected intensity. But to investigate the magnitude of the effect quantitatively, it becomes necessary to adopt a version of the reflectivity expression that accounts for the extinction of the incident wave as it progressively encounters more planes on penetrating deeper into the crystal. A good approximation for such a so-called "dynamical" reflectivity $|r_{DYN} (Q_Z)|^2$ follows from the expression given by Equation B 1 as derived, for example, by Zachariasen [35]

(B 4)     $|r_{DYN} (Q_Z)|^2 = \tanh^2 \{ (4\pi / Q_Z) \rho_{PG(002)} [\sin(NQ_Z d/2) / \sin(Q_Z d/2)] \}$

At the maximum where $Q_{Z\,MAX} = 2\pi / d$ , Equation B 4 reduces to

(B 5)     $|r_{DYN} (Q_{Z\,MAX} = 2\pi / d)|^2 = \tanh^2 \{2dN \rho_{PG(002)} \}$

It can then be asked what number N of atomic planes contributes to a reflectivity of, say, $|r_{DYN} (Q_{Z\,MAX} = 2\pi / d)|^2 = 0.95$. For a perfect crystal mosaic block of HOPG, $\rho_{PG(002)} = 2.54 \times 10^{-5}$ Å$^{-1}$ , $2d = 6.7078$ Å and N is found to be about 12,807 planes (or a crystal block thickness of about 4.3 microns). For comparison, another common monochromator is perfect single crystal silicon -- for the Si(111) reflection, $\rho_{Si(111)} = 3.25 \times 10^{-6}$ Å$^{-1}$ (almost an order of magnitude less than that for $\rho_{PG(002)}$ ), d = 3.135 and N turns out to be approximately 106,900 planes (a crystal thickness of about 33.5 microns) to attain a reflectivity of 0.95.

This information may be of interest in designing neutron optical elements for diffractometers and ultra small angle scattering (USANS) instruments (see, for example, [19], [14]), but to be of further use in helping to understand the transverse coherent extent of a packet wavefront, it is necessary to go beyond a description of the diffraction based on single plane waves of infinite lateral extent. That is, a theory for the scattering of wave packets that is intrinsically time-dependent is required for a rigorous treatment (see, for instance, [16]). Nonetheless, for present purposes, an approximation of wave packet scattering can be made based on the 2D wave train function employed in other sections of the main body of the paper in which the magnitude of k for all plane wave components is nearly constant. In this way, a superposition of wave amplitudes analogous in spirit to the Huygens-Fresnel construction applied to circular waves emanating from an aperture can be applied.

As in the Huygens-Fresnel construction applied to the single aperture earlier, we will neglect multiple scattering effects and use the kinematic expression for the reflection amplitude $r_{BA}(Q_Z)$, the corresponding reflectivity for which is given in Equation B 4. In place of summing over all of the spherical (circular in two dimensions) wavelets emanating from each source point, we first calculate the reflected wave amplitudes from the N atomic planes of the crystalline block which result for an incident plane wave at a specified angle of incidence corresponding to each of the wavevector values in the distribution of plane wave momentum basis states that compose an incident neutron 2D wave train. We assume a Gaussian distribution of the basis state wavevectors and weight each of these contributions accordingly before summing over all of the individual reflection amplitudes to obtain a reflected wave with a mean wavevector $\mathbf{k_M}$. The reflection amplitude $r_{BA}(Q_Z = 2k_Z$ ) for each plane wave component has the form

(B 6)     $r_{BA}(k_Z) = [4\pi / (i2k_Z)] \{\exp[+ik_Z (N-1)d]\} \rho_{PG(002)} [\sin(Nk_Z d) / \sin(k_Z d)]$

Figure B 3 is a schematic of the reflection geometry. In this diagram, the reference frame of the incident wave is shown in addition to the laboratory or crystalline block reference frame. The $z\perp$ and $x\|$ orthogonal axes define the neutron frame whereas $z_L$ and $x_L$ correspond to the laboratory frame. $\mathbf{k_{Ij}}$ labels the jth basis wavevector component of the incident wave train distribution that is incident at an angle $\theta_j$ relative to the surface atomic plane of the crystal. $\mathbf{k_{IM}}$ , on the other hand, labels the mean or central wavevector of the wave train. The corresponding scattered or reflected wavevectors are labeled with the subscript "s" in a respective manner.

For the remainder of this discussion, the special 2D elongated wave train function that has been adopted to represent the neutron packet will be referred to in more generic terms simply as a wave packet. Since the primary interest is in how reflection from a crystal might affect the transverse distribution of packet wavevector components perpendicular to the mean packet wavevector $\mathbf{k_M}$ , it would be convenient to select the constituent basis wavevectors of the packet to all have the same longitudinal components along the $x\|$ axis of the packet reference frame. However, from the discussion in the section on diffraction from a single slit aperture in the main body of text regarding allowable solutions for the time-independent wave equation, such a solution requires that $k^2 = k^2_X + k^2_Y + k^2_Z$ = a constant for all the basis function wavevectors. Nevertheless, for the relatively narrow spread of packet basis wavevectors of relevance here, it turns out that the variation in the longitudinal components given the condition that k be a constant value is negligible. Recall the example given in the earlier discussion. For an angular range of basis wavevector orientations of the order of $\varepsilon = 5$ arc seconds ($2.424 \times 10^{-5}$ radians) about the mean wavevector direction, the constraint that $k^2$ be a constant value (say $k = 2\pi / 5$ Å) results in a magnitude variation of the longitudinal wavevector component of approximately $k [ 1 - \cos (\varepsilon) ] = 2.94 \times 10^{-10}$ k whereas the transverse (perpendicular) component variation is $k \sin(\varepsilon) = 4.23 \times 10^{-7}$ k (or 1439 times larger). (As should become clear in the derivation that follows, the longitudinal components in principle can also contribute to the transverse width of the packet upon reflection, but this effect will be neglected here for simplicity since it does not affect the fundamental idea in any essential way.)

Note that $k\perp$ as referred to the neutron packet reference frame has a projection along the z-axis of the crystal frame related by (see Figure B 3)

(B 7)                                   $k\perp = (k_{MZ} - k_Z) / \cos \theta_M$

The relation immediately above can be used to write the distribution of transverse wavevector components for the incident packet (taken to be Gaussian), W( $k_Z$ ) in terms of the variable $k_Z$ in the crystal frame as

(B 8)        $W(k_Z) = (1 / \Gamma_{FWHM\,k\perp}) (4 \ln2 / \pi)^{1/2} \exp\{-[(k_{MZ} - k_Z) / \cos \theta_M]^2 / [(\Gamma_{FWHM\,k\perp})^2 / (4 \ln2)]\}$

where $\Gamma_{FWHM\,k\perp}$ is the width (FWHM) of the transverse wavevector component distribution for the incident packet and where $k_{MZ}$ and $\cos \theta_M$ are constant values which correspond to the z-component of the mean wavevector and angle of incidence, respectively, of a single neutron packet .

Thus even though the component of the wavevector along the crystal x-axis (in the atomic reflecting plane) remains constant in the specular reflection process, the other component along the wavevector transfer direction parallel to the crystal z-axis can be affected.  The consequence of this is that the distribution of transverse wavevector components in the outgoing reflected neutron packet can be altered from that in the incident packet in the course of the selective Bragg diffraction process.

The reflection amplitude for a packet $r_{WP}(k_{MZ})$ is then represented, in this approximation, by a weighted summation of reflection amplitudes for each of the packet's component basis states using Equation B 6 (and writing the atomic plane scattering length density for a generic case as $\rho_{AP}$):

(B 9)  $r_{WP}(k_{MZ}) = \int r_{BA}(k_z)\, dk_z = \int [4\pi / (i2k_z)]\, W(k_z)\, \{\exp[+ik_z\,(N-1)d]\}\, \rho_{AP}\, [\sin(Nk_zd) / \sin(k_zd)]\, dk_z$

where $W(k_z)$ is given by Equation B 8.  The integration in Equation B 9 is over the distribution represented by W.  The sine functions in Equation B 9 modulate the Gaussian distribution weighting, depending upon the relative values of $\Gamma_{FWHM\,k\perp}$ and N, narrowing the width of the distribution of transverse wavevector components in the reflected packet in certain cases -- and, consequently, resulting in a broader width of the packet $\Delta r_T$ in real space.

However, as discussed previously, a complete picture requires that a distribution of packet mean wavevectors corresponding to the ensemble of neutrons composing a beam be taken into account as well.  So, in addition to the coherent superposition of amplitude components for an individual packet, an incoherent averaging over the mean wavevectors of the packets in the collection making up the beam is necessary.  As before, for simplicity, we consider the special case in which all of the packets are of identical size and shape but have different mean wavevector *directions* which characterize an angular distribution or beam geometrical angular divergence.

The geometrical angular divergence of such a beam, as described above, can be associated with a Gaussian distribution $W_{BEAM}\,(\theta_M)$ (each angle $\theta_M$ corresponding to a $k_M$ )

(B 10)    $W_{BEAM}\,(\theta_M) = (1 / \Gamma_{FWHM\,\theta M})\,(4\ln2 / \pi)^{1/2}\, \exp\{-[<\theta_M> - \theta_M]^2 / [(\Gamma_{FWHM\,\theta M})^2 / (4\ln2)]\}$

where $< \theta_M >$ denotes the average value of wave packet angle in the ensemble of packets composing the beam.  The intensity distribution in the reflected beam $I_{RB}\,(\theta_M)$ as a function of angle is then given by

(B 11)                    $I_{RB}\,(\theta_M) = |r_{WP}(k_{MZ})|^2\, W_{BEAM}\,(\theta_M)$

where $k_{MZ} = k_M \sin(\theta_M)$.  Note that $W_{BEAM}\,(\theta_M)$ describes the incident beam angular distribution whereas $I_{RB}\,(\theta_M)$ refers to the reflected beam intensity distribution.

In Figure B 4 are plotted examples of the incident and reflected beam intensities as a function of the transverse component of packet mean wavevector for a Si(111) monochromator crystal as calculated according to the approximate theory introduced above. The transverse component of packet mean wavevector $k_{MT} = k_M \tan(\alpha)$ -- where $\alpha$ is the angle of deviation a given neutron packet's mean wavevector $k_M$ makes relative to the average packet mean wavevector $<k_M>$ in a geometrically collimated beam. The examples correspond to crystal blocks with either N equal to $10^4$ or $10^5$ atomic planes and for an incident beam divergence of $1.4 \times 10^{-4}$ radians (FWHM) and a packet transverse wavevector component distribution width $\Gamma_{FWHM\ k\perp}$ ( $= \Delta k_T$) of $1.76 \times 10^{-4}$ Å$^{-1}$. Reflection from the crystal narrows the angular divergence of the beam through the Bragg diffraction process but also simultaneously reduces the width of the transverse wavevector component distribution intrinsic to each individual reflected wave packet as shown in the lower half of the figure -- as discussed earlier, the narrowing is inversely proportional to the number of atomic planes contributing to the coherent reflection process. The narrowing of the reflected packet transverse wavevector component distribution becomes more pronounced with increasing width of the incident packet's distribution. If the incident distribution is already relatively narrow, the width of the reflected distribution may not be significantly reduced in comparison.

What has been described immediately above corresponds to a relatively simple, limiting case where one neutron interacts with a single mosaic block -- other more complicated situations may arise where combinations of mosaic blocks are simultaneously illuminated by the same incident neutron. Also, as already mentioned, multiple scattering effects have not been taken into account. Nonetheless, the simplified approximate treatment captures the essence of how the Bragg reflection process from a perfect crystal mosaic block of finite size can define the transverse extent of the wavefronts of a reflected wave packet.

Appendix C: Mutual Coherence Function -- *Computational Details*

As discussed previously in the section on the mutual coherence function in the main body of the paper, to examine the relation between fringe visibility and the pertinent characteristics of the radiation source, it is necessary only to evaluate the informational content of the normalized mutual coherence function or complex degree of coherence $\gamma_{A1A2}(\tau)$ with the explicit form for the neutron wave function $\Psi(r,t)$ taken to be effectively planar at the pertinent location.

The relevant part of Figure 8 is redrawn in more detail in Figure C1. Rather than compute the time average associated with the mutual coherence function as discussed in the main section on the mutual coherence function, we will invoke ergodicity and perform the equivalent, alternative ensemble average over all possible realizations of the neutron waves emanating from the extended source.

Let the wave field emanating from each point on the extended source be circular and far enough from the apertures that the curvature of every wave front at the location of the apertures can be approximated by a straight line (plane-wave-like in the Fraunhofer limit). The wave function in this (two-dimensional case) is given by

(C1) $$\Psi(r) = \Psi(x,y) = [1/(2\pi)] \exp[i(k_x x + k_y y)]$$

where $k_x$ and $k_y$ are the Cartesian components of the neutron wavevector.  The quantity that we need to calculate explicitly is

(C2) $\gamma_{A1A2} = < \Psi_{A1} (x_1=0,y_1=+a) \; \Psi_{A2}^{*} (x_2=0,y_2=-a) > / [<|\Psi_{A1} (x_1=0,y_1=+a)|^2> <|\Psi_{A2} (x_2=0,y_2=-a)|^2>]^{1/2}$

where the ensemble average <> is carried out over the entire line source from y = -s to y = +s.  It is assumed that the source emits a uniform distribution of waves at all positions along the line (parallel to the y-axis).  The neutron wavevector direction is specified by the angles $\theta_1$ and $\theta_2$ for positions A1 and A2 on the y-axis at +a and -a -- where the wave functions are evaluated -- as indicated in Figure C1.  The parameters $r_1$ and $r_2$ specify the corresponding distances between a particular source point on the line and the points A1 and A2.  An integration along the source line can be straightforwardly parameterized by expressing the x- and y-components of the wavevector, $k_x$ and $k_y$, in terms of $\theta_1$, $\theta_2$, l and a.  Note that the plane wave fronts are approximated to be perpendicular to the line between the origin and the source point -- which should be valid for the relatively small angles typically encountered.  The essential requirement is to properly account for the phase difference between the two points A1 and A2.   We can write

(C3)        $\Psi_{A1} (x_1=0,y_1=+a) \; \Psi_{A2}^{*} (x_2=0,y_2=-a) = [1/(2\pi)^2] \exp[i(k_{x1}x_1 + k_{y1}y_1)] \exp[-i(k_{x2}x_2 + k_{y2}y_2)]$

$= [1/(2\pi)^2] \exp[-i(ka/l)(2Y)]$

where we have made the substitutions $k_{y1} = k \sin\theta_1$, $k_{y2} = k \sin\theta_2$, $\tan\theta_1 = [(Y-a)/l] \approx \sin\theta_1 \approx \theta_1$, $\tan\theta_2 = [(Y+a)/l] \approx \sin\theta_2 \approx \theta_2$, and where Y has been defined as the position along the vertical line source axis.  Then

(C4)        $< \Psi_{A1} (x_1=0,y_1=+a) \; \Psi_{A2}^{*} (x_2=0,y_2=-a) > = [1/(2\pi)^2] \; _{-s}\!\int^{+s} \exp[-i(ka/l)(2Y)] \; dY$

$= [2s/(2\pi)^2] \sin[(2s)(2a)\pi/(l\lambda)] / [(2s)(2a)\pi/(l\lambda)]$

where $k = 2\pi/\lambda$ has been substituted in which $\lambda$ is the neutron wavelength.  Performing similar integrations we obtain

(C5)        $<|\Psi_{A1} (x_1=0,y_1=+a)|^2> = <|\Psi_{A2} (x_2=0,y_2=-a)|^2> = [2s/(2\pi)^2]$

so that

(C6)     $\gamma_{A1A2} = \sin[(2s)(2a)\pi/(l\lambda)] / [(2s)(2a)\pi/(l\lambda)] = \operatorname{sinc}[(2s)(2a)\pi/(l\lambda)]$

which is a well-known result in classical light optics.  As previously noted, the modulus of $\gamma_{A1A2}$ is equal to the fringe visibility given by Equation 5 (for the case of the two-slit interference pattern).

The implications of this result are discussed in the section on the mutual coherence function in the main body of the paper.

Appendix D: Instrumental Specifications

The generic neutron reflectometer or diffractometer schematically illustrated in Figure 1 is more quantitatively described here.  At the NIST Center for Neutron Research (NCNR), for instance, a typical reflectometer is configured as follows.  An HOPG (Highly Oriented Pyrolytic Graphite) monochromator situated in a gap on the NG-D neutron guide tube is aligned to Bragg reflect neutrons with wavelengths of 5.00 +/- 0.01 Å (deviation FWHM) (the wavelength range being partially determined by the angular collimations of the beams incident and reflected).  The HOPG monochromator is referred to as ideally imperfect since it is composed of single crystal "mosaic" blocks, each of several micron dimensions, that have a preferred orientation of the (002) atomic planes of the hexagonal close-packed structure of pyrolytic graphite parallel to the nominal surface of the crystal but are randomly oriented in-plane.  The action of such a monochromating device for neutrons is discussed further in Appendix B.  It is expected that for typical HOPG monochromators the reflected neutrons have wave packets with transverse dimensions of the order of tens of microns at a distance of a meter or so away (this is consistent with the diffraction measurements described here and in other works [19]).  The distance between the monochromator and the first slit aperture downstream is approximately 908. mm (in between there is a polycrystalline block of Be cooled to liquid nitrogen temperature to remove higher-order wavelength neutrons).  The distance between the first and the following second slit is L = 863.6 mm.  Both slits have a nominal width W = 0.025 mm. Each slit aperture extends along the vertical direction -- out of the plane of Figure 1 -- by approximately 25. mm and is defined by a pair of parallel absorbing Cd masks 1 mm thick with accurately machined edges. An identical "detector" slit is situated in front of a $^3$He detector tube a distance of 1651. mm away from the second slit.  The detector slit can be scanned along a direction perpendicular to this axis defined along the center of the two collimating slits upstream.  The geometrical divergence limits +/- $\varepsilon$ of the pair of collimating slits is given by $\tan(\varepsilon) = W / L$.  For the numerical values given above, $\varepsilon$ = +/- 2.895 x $10^{-5}$ radians 1.659 x $10^{-3}$ degrees = 9.952 x $10^{-2}$ minutes of arc = 5.97 seconds of arc).  Projecting back through the pair of slits toward the HOPG monochromator upstream,  the perpendicular width viewed at the monochromator position is approximately 7.76 x $10^{-2}$ mm (although a width of about 25. mm across the surface of the monochromator illuminates the entrance to the first slit with a relatively uniform -- spatially and in angle -- flux of quasi-monochromatic neutrons).

References

[1] N.J. Chesser and J. D. Axe, *Acta Cryst.* (1973), **A29**, 160, "Derivation and Experimental Verification of the Normalized Resolution Function for Inelastic Neutron Scattering".

[2] Ronald L. Cappelletti, Terrence Jach, and John Vinson, *Physical Review Letters* **120**, 090402 (2018), "Intrinsic Orbital Angular Momentum States of Neutrons".

[3] P.A.M. Dirac, *Quantum Mechanics, 4th Ed.*, (Oxford University Press, London, 1958) p.9.

[4] Majkrzak, Charles F., Metting, Christopher,Maranville, Brian B., Dura, Joseph A., Satija, Sushil, Udovic, Terrence, Berk, Norman F., *Physical Review* **A 89**, (2014) 033851, "Determination of the effective transverse coherence of the neutron wave packet as employed in reflectivity investigations of condensed-matter structures. I. Measurements".

[5] A. Born and E. Wolf, *Principles of Optics*, (Pergamon Press, Oxford, 1975).

[6] L. Mandel and E. Wolf, *Optical Coherence and Quantum Optics*, (Cambridge University Press, Cambridge, 1995); L. Mandel and E. Wolf, *Reviews of Modern Physics* **37**, (1965) 231, "Coherence Properties of Optical Fields".

[7] E. Hecht, *Optics, 3rd Ed.*, (Addison Wesley, New York, 1998).

[8] R. Gaehler, J. Felber, F. Mezei, and R. Golub, *Phys. Rev.* **A 58**, 280 (1998).

[9] J. Felber, R. Gaehler, R. Golub, and K. Prechtel, *Physica* **B 252**, 34 (1998).

[10] V.O. de Haan, A.A. van Well, and J. Plomp, *Phys. Rev.* **B 77**, 104121 (2008); V.O. de Haan, J. Plomp, M.Y. Rekveldt, A.A. van Well, R.M. Dagliesh, S. Langridge, A.J. Boettger, and R. Hendrikx, *Phys. Rev.* **B 81**, 94112 (2010).

[11] S.K. Sinha, M. Tolan, and A. Gibaud, *Phys. Rev.* **B 57**, 2740 (1998).

[12] M. Tolan, G. Koenig, L. Bruegemann, W. Press, F. Brinkop, and J.P. Kotthaus, *Europhys. Lett.* **20**, 223 (1992); M. Tolan, W. Press, F. Brinkop, and J.P. Kotthaus, *J. Appl. Phys.* **75**, 7761 (1994).

[13] T. Salditt, H. Rhan, T.H. Metzger, J. Peisl, R. Schuster, and J.P. Kotthaus, *Zeitshrift fur Physik B* **96**, 227 (1994).

[14] V.M. Kaganer, B. Jenichen, and K.H. Ploog, *J. Phys. D: Appl. Phys.* **34**, (2001) 645-659, "Crystal optics elements in a coherent x-ray scattering experiment".

[15] J. Stoehr, *Synchrotron Radiation News* **32**, (2019) 48-51.

[16] N.F. Berk, arXiv:1811.06054, 14 Nov 2018 20:48:34, "On Scattering of 1-D Wave Packets".

[17] K.Y. Bliokh, I.P. Ivanov, G. Guzzinati, L. Clark, R. Van Boxem, A. Beche, R. Juchtmans, M.A. Alonso, P. Schattschneider, F. Nori, and J. Verbeeck, "Theory and applications of free-electron vortex states", *Physics Reports* **690**, (2017) 1-70.

[18] A. Zeilinger, R. Gaehler, C.G. Shull, W. Treimer, and W. Mampe, *Rev. Mod. Phys.* **60**, 1067 (1988), "Single- and double-slit diffraction of neutrons".

[19] W. Treimer, A. Hilger, M. Stroble, *Physica B* **385-386**, (2006) 1388-1391, "Slit and phase grating diffraction with a double crystal diffractometer".

[20] S.W. Lee, D.S. Hussey, D.L. Jacobson, C.M. Sim, and M. Arif, *Nucl. Instrm. and Meths. in Phys. Res. A* **605** (2009) 16-20, "Development of the grating phase neutron interferometer at a monochromatic beam line".

[21] R.L. Pfleegor and L. Mandel, *Phys. Rev.* **159**, (1967) 1084-1088; "Interference of Independent Photon Beams".

[22] H.-J. Lee, C.L. Soles, and W.-l. Wu, *ECS Transactions* **34**, 931 (2011).

[23] T. Salditt, H. Rhan, T.H. Metzger, J. Peisl, R. Schuster, and J.P. Kotthaus, *Zeitshrift fur Physik* B **96**, 227 (1994).

[24] M. Tolan, G. Koenig, L. Bruegemann, W. Press, F. Brinkop, and J.P. Kotthaus, *Europhys. Lett.* **20**, 223 (1992); M. Tolan, W. Press, F. Brinkop, and J.P. Kotthaus, *J. Appl. Phys.* **75**, 7761 (1994).

[25] L.E. Ballentine, *Quantum Mechanics, a Modern Development*, (World Scientific, Singapore, 1988).

[26] H. Rauch and S. Werner, *Neutron Interferometry*, 2nd Ed. (Clarendon Press, Oxford, 2015).

[27] V.F. Sears, *Neutron Optics*, (Oxford University Press, Oxford, 1989).

[28] A. Tonomora, J. Endo, T. Matsuda, T. Kawasaki, and H. Exawa, "Demonstration of single-electron buildup of an interference pattern" *Amer. J. Phys.* **57**, (1989) p. 117.

[29] T. Maudlin, *Part and Whole in Quantum Mechanics*, p. 52, in E. Castellani, Edt., *Interpreting Bodies: Classical and Quantum Objects in Modern Physics*, Princeton University Press, Princeton, New Jersey, (1998).

[30] Alexia Auffees and Philippe Grangier, *Contexts, Systems, and Modalities: a new Ontology for Quantum Mechanics*, arXiv:1409.2120v [quant-ph] 18 Apr 2009.

[31] E. Merzbacher, Quantum Mechanics, 1961, John Wiley & Sons (New York), Chapter 2, pp. 22.

[32] W. Marshall and S.W. Lovesey, Theory of Thermal Neutron Scattering, Oxford Clarendon Press (1971); Introduction to the Theory of Thermal Neutron Scattering, Cambridge University press, 3rd Edition, 2012.

[33] G.C. Ghirardi, A. Rimini, and T. Weber, "Unified dynamics for microscopic and macroscopic systems", *Phys. Rev. D* **34**, 470 (1986); A. Bassi, S. Satin, T. Singh, and H. Ulbricht, "Models of wave-function collapse, underlying theories, and experimental tests", *Rev. Mod. Phys.* **85**, 476 (2013).

[34] M. Utsuro and V.K. Ignatovich, *Handbook of Neutron Optics*, (Wiley-VCH, Weinheim, 2010).

[35] W.H. Zachariasen, *Theory of X-Ray Diffraction in Crystals*, Wiley, New York, 1945.

Figures and Tables

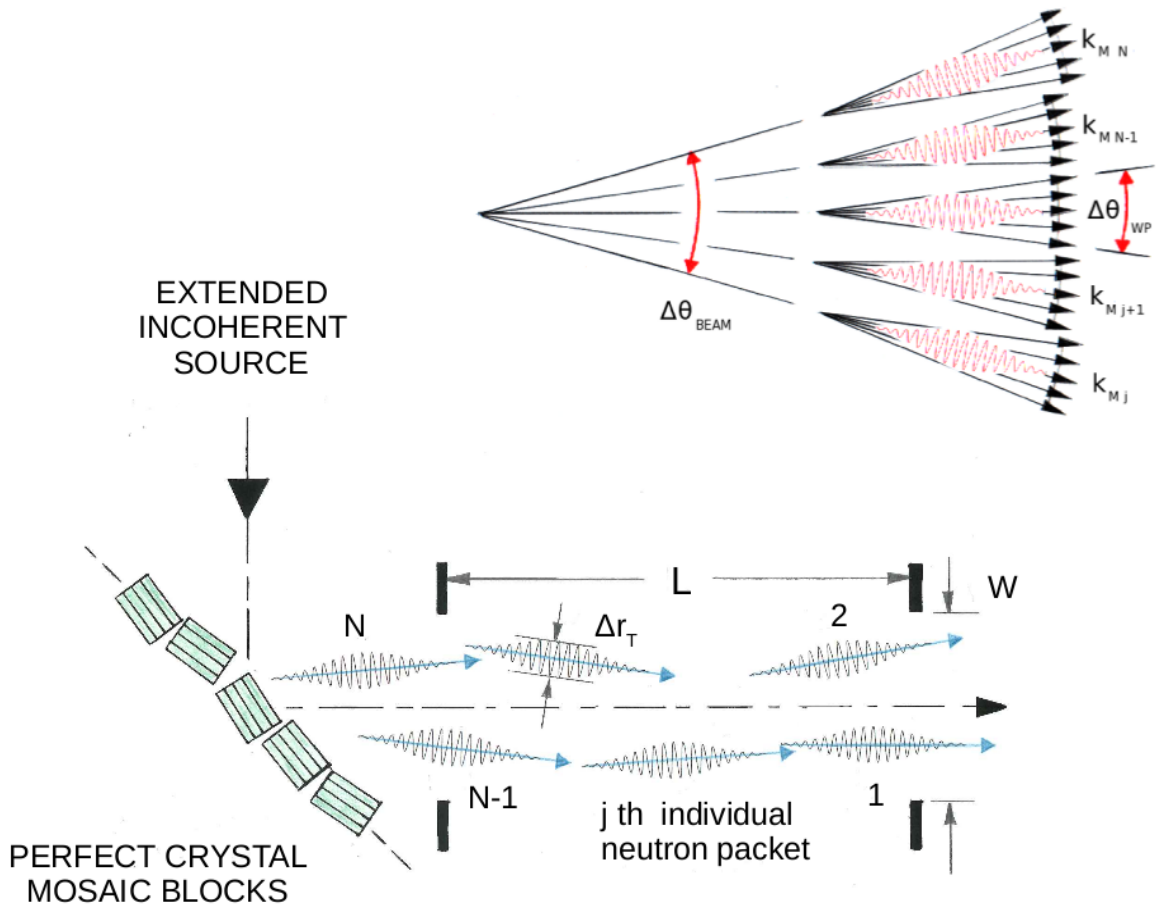

Figure 1. Schematic representation of the essential elements of a rudimentary neutron diffractometer just upstream of the sample position (which would be located to the right of the beam exit slit). An ideally imperfect mosaic crystal (e.g., pyrolytic graphite) directs incident neutrons (by energy-selective Bragg reflection), originating from a temporally and spatially extended incoherent source, through a pair of slits resulting in a quasi-monochromatic beam to be incident on a sample. This beam is a collection of individual neutrons, each with an associated wave packet. The j th individual wave packet corresponds to one specific neutron that is a member of a collection of similar neutrons. Each packet has a mean wavevector $k_M$. The beam is further characterized by a distribution of packet mean wavevectors that define a geometrical angular divergence related to W/L. But as shown in the inset at the top, each packet is itself composed of a coherent distribution of basis states with corresponding eigenvectors. The resultant picture is one in which both coherent and incoherent distributions of wavevectors corresponding to individual neutron packet and beam, respectively, coexist. The monochromator and pair of apertures together define both the individual and collective properties of the packets and beam, respectively, as described in the main text.

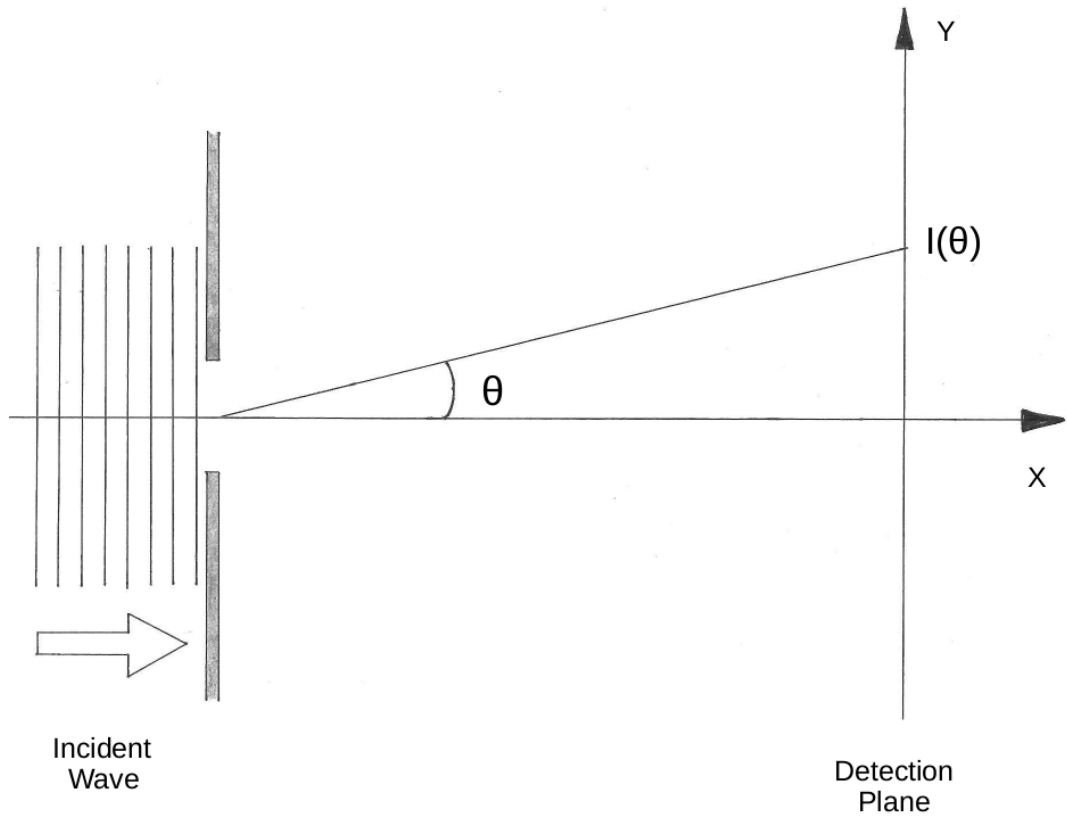

*Figure 2. Schematic for single slit diffraction in two dimensions assuming a monochromatic wave incident from the left. The incident plane wave from the left is schematically represented with truncated wavefronts.*

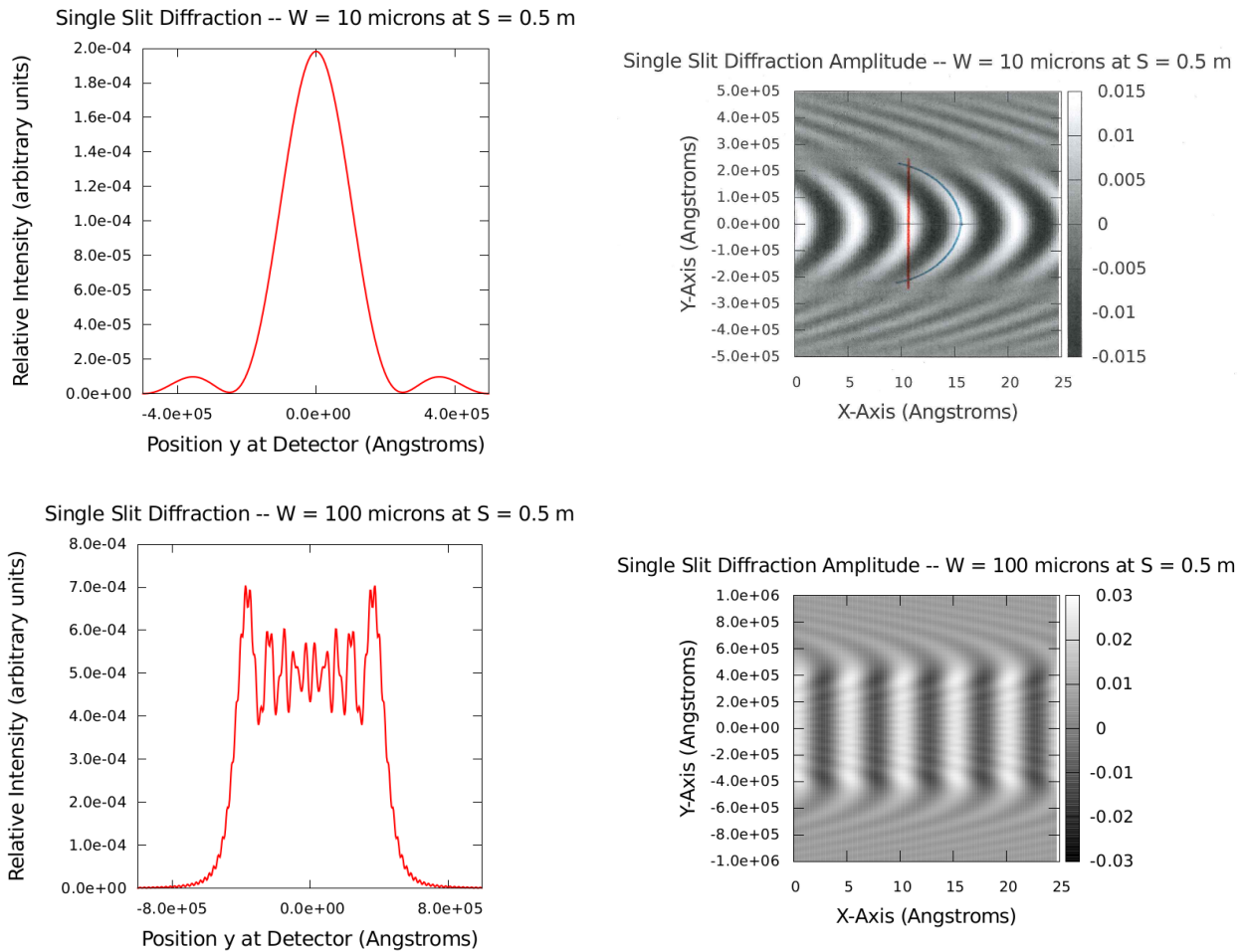

*Figure 3. Intensity patterns (left-hand column) and the real parts of the diffraction amplitudes (right-hand column) from slit apertures (for the two-dimensional geometry shown in Figure 2) corresponding to two of the examples listed in Table 2. The intensity patterns are plotted at a point of observation a distance S (0.5 m) away from the aperture along the y-axis perpendicular to the direction of propagation of the wave incident on the slit. For the amplitude plots, the y-axis is also perpendicular to the incident beam direction whereas the x-direction extends a distance of five neutron wavelengths along the direction of propagation <u>back</u> from the farthest observation point at a distance S. One measure of the distance over which a given wavefront is uniform in phase to within one wavelength can be determined by examining two consecutive wavefronts propagating along the x-axis as pictured in the upper right-hand plot. Identify the point of maximum amplitude on the ridge of the leading wavefront (shown as a blue curve) that has the same transverse y-value as that of a point on the ridge of the following front (the intersection of the blue curve and red line). The x-coordinates of these two points differ by λ. For the 10 micron slit, this measure corresponds roughly to that obtained from the first minimum of the intensity on either side of the central maximum. The 10 micron wide slit produces a pattern at 0.5 m that is in the far-field limit while the 100 micron aperture is well within the near-field region. Considering the amplitude waveforms on the right, it can be seen that the wavefronts propagating outward from the aperture have a well-defined finite lateral extent (although that extent increases with distance).*

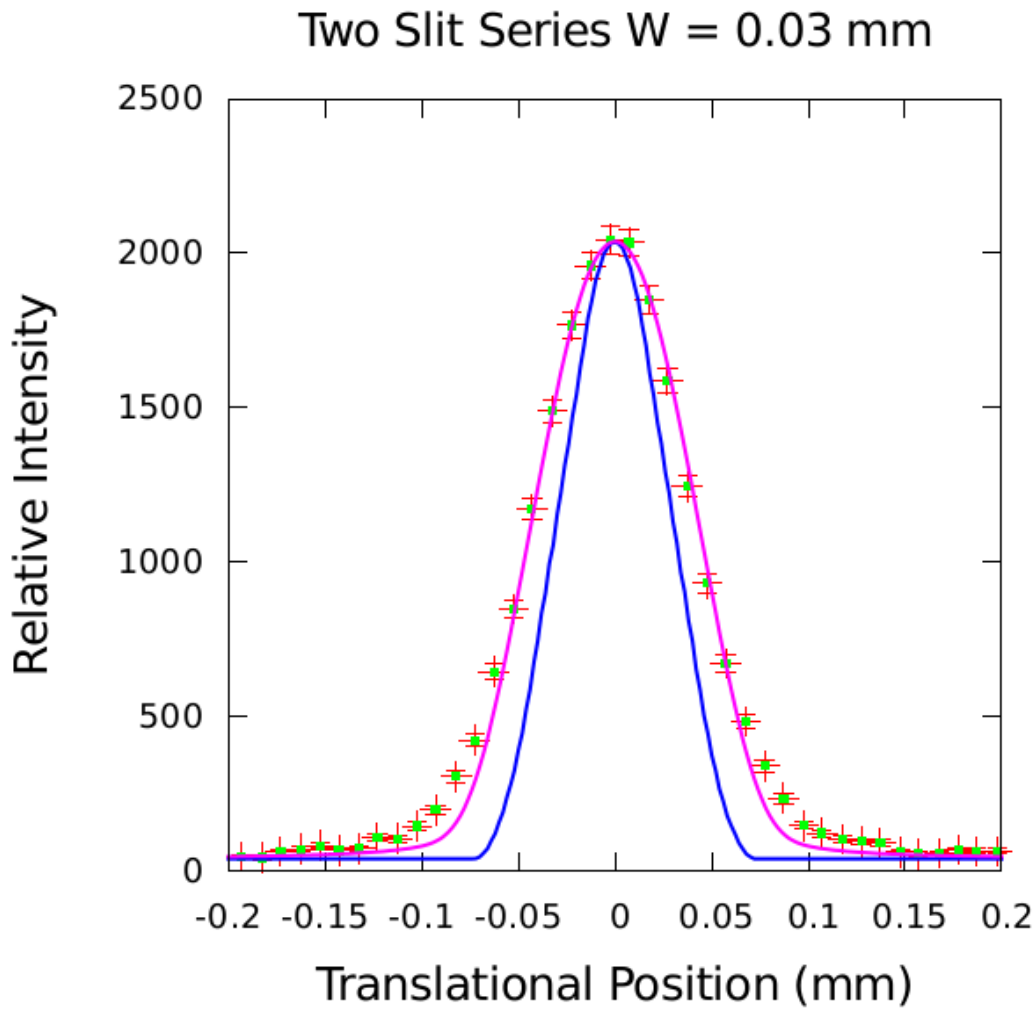

*Figure 4. Numerical calculation (purple curve) of the beam profile expected to be projected onto the detector line on an instrument as compared to an actual measurement (points). The computed curve is not a fit, but only scaled to the measured intensity as described in the main text. In this calculation, the geometrical angular limits defined by the pair of slits together were taken into account and a summation of diffracted intensity patterns from the second slit downstream -- as predicted by the standard Fraunhofer diffraction formula for multiple source points across the width of the first (upstream) slit -- was performed. Another calculation, corresponding to what would be expected based on a consideration of geometrical ray optics alone (blue curve) is also plotted. The agreement in the case in which both geometrical and diffraction effects are included is markedly better.*

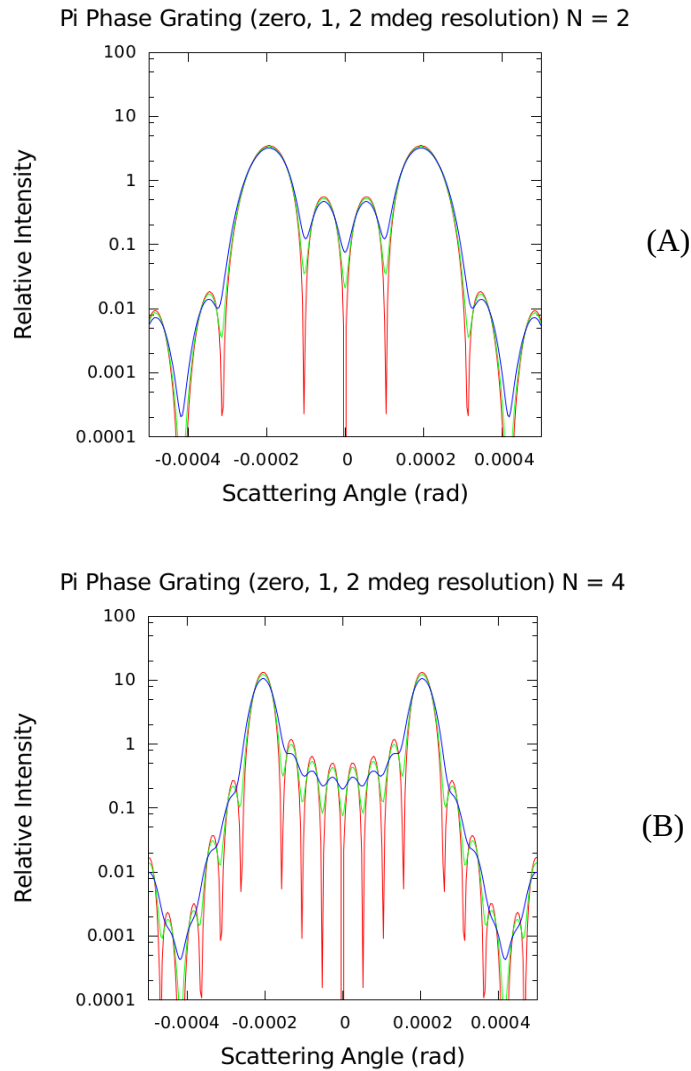

*Figure 5. Diffraction patterns calculated for a model π phase-shift grating with equal column and trough thicknesses at a 5 Å neutron wavelength. One set of patterns corresponds to coherent contributions from 2 grating periods (A) and the other set from 4 (B). The period of the grating is 2.4 microns. Assuming the grating itself is perfectly uniform, the number of coherently contributing periods then depends on the transverse width of the neutron wavefront over which the phase is of the requisite uniformity. The geometrical angular divergence of the incident neutron beam -- which corresponds to a distribution of transverse components of packet mean wavevectors -- determines how well the features of the pattern are resolved. Both figures include cases for zero beam divergence and two other finite values (3.6 and 7.2 sec of arc) convoluted with the natural pattern. Note that the general shape of the pattern is preserved while the widths and magnitudes of the principle and subsidiary reflections are affected by the degree of the geometrical angular divergence of the incident beam. This description in which a distinction can be made between the effect of the intrinsic transverse width of an individual neutron packet and that of the beam geometrical angular divergence is supported by measurements reported here (which follow) and those of others [19].*

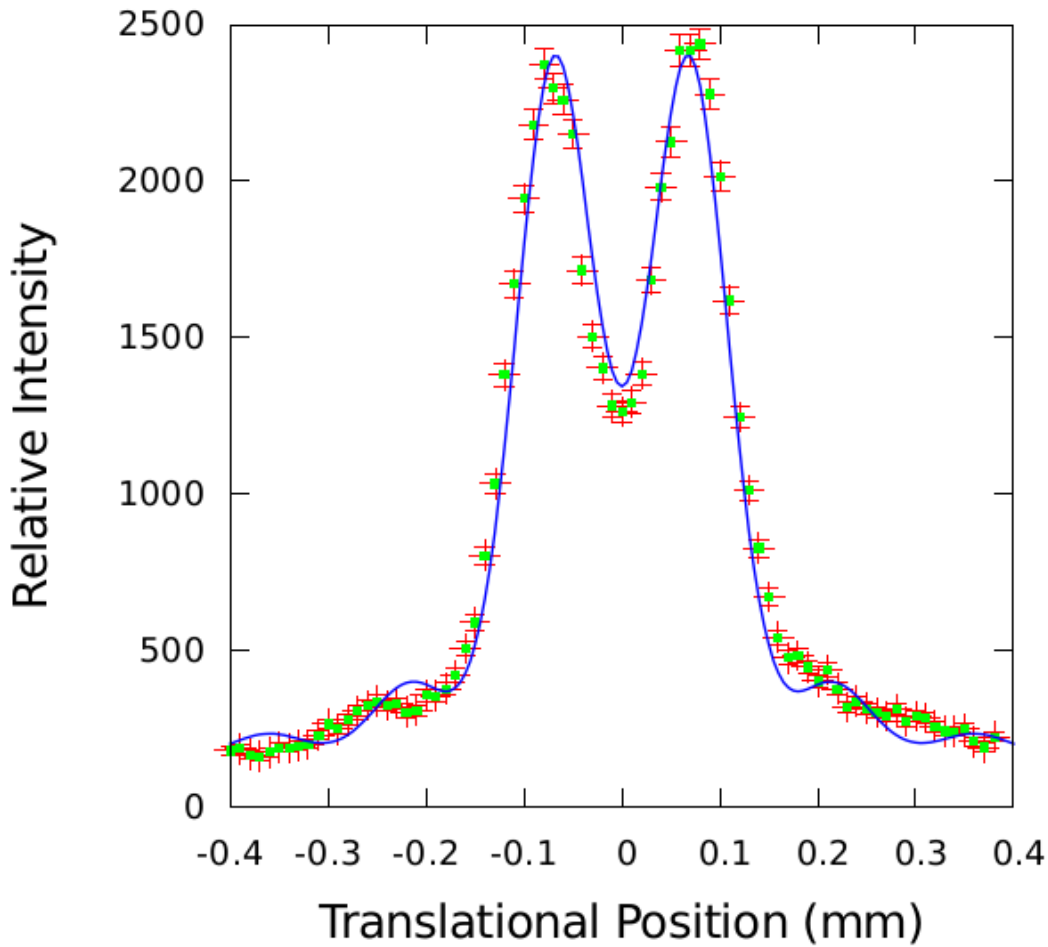

Figure 6. Diffraction pattern measured (symbols with error bars) from an 8 micron period π phase-shift grating with equal-thickness rectangular troughs and columns etched in single crystal silicon at a neutron nominal wavelength of 5 Å. Also plotted in this figure is a calculated diffraction pattern based on the phase grating formula of Equation 4 and assuming an incident illuminating beam described exactly as in the preceding section for the pair of slits which resulted in the profile of Figure 4. That is, both geometrical and diffraction effects in forming the beam incident on the grating by the pair of slits were taken into account. In the computed phase grating diffraction pattern, both the distribution of geometrical angles in the incident beam and the transverse dimension of an individual neutron packet wavefront were included. The model calculation was not fit to the data, but only scaled to the measured intensity. The best agreement between the data and model was obtained for N = 3 and for a slight curvature of the grating substrate amounting to about 2.65 x 10⁻⁵ radians (5.47 seconds). (This bending might alternatively be attributed to a curvature of a neutron packet wavefront -- which was originally taken to be perfectly flat but limited to a 24. micron finite lateral extent.) The general agreement between measurement and model calculation is good, although details in the wings are not resolved -- this is likely due to relatively small effects involving mirror reflection, refraction, and diffraction from the mask edges of the slits defining the incident beam.

A) Totally coherent multiple (aperture) sources simultaneously illuminated by a single (planar) incident wavefront.

B) Completely incoherent multiple sources illuminated by multiple, independent incident packet wavefronts, each of finite transverse extent capable of spanning no more than a single aperture width -- the incident beam intensity is low enough that only one aperture is illuminated at a time, i.e., there is negligible overlap between successive packets in the beam.

C) Predominantly incoherent multiple sources illuminated by independent incident wave packets as in case B above but where the beam intensity is sufficiently high that some incident wave packets are close enough in time and phase relative to one another that a partial, coincidental correlation is possible between two or more of the secondary source apertures.

D) A partially coherent source where, for example, two adjacent source apertures are simultaneously illuminated by the same incident packet wavefront.

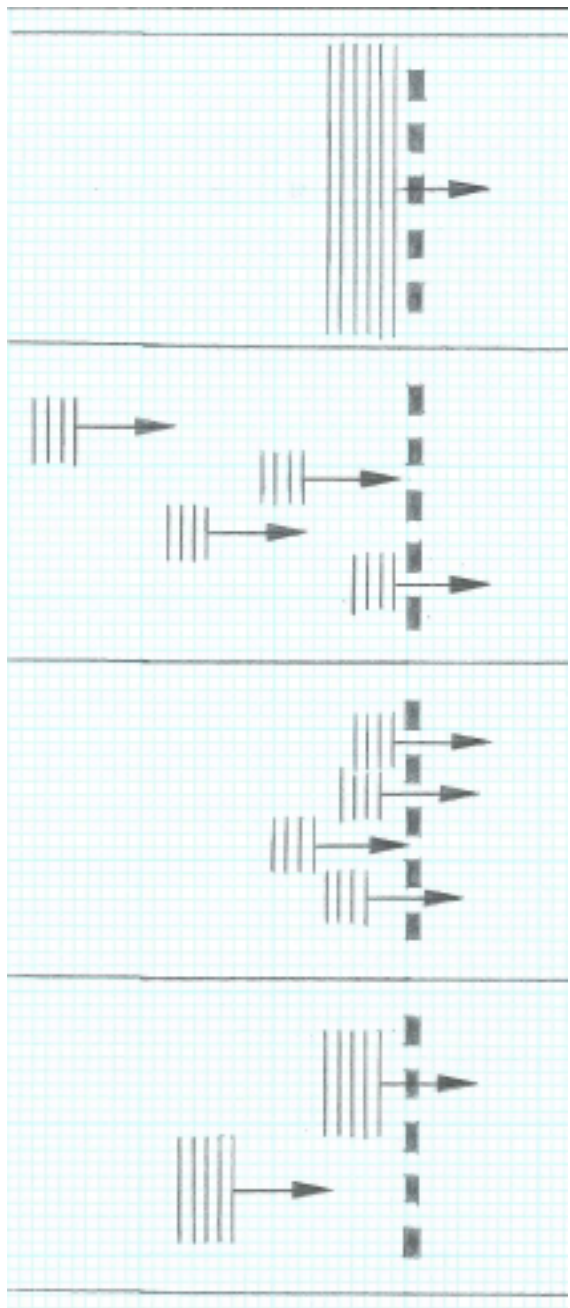

*Figure 7.  Extended line sources represented by multiple apertures.  The coherence of the radiation emitted from each secondary source example depends upon the nature of the radiation incident -- represented schematically as packets with truncated planar wavefronts -- on the aperture array from a primary source to the left.  A number of possible cases, A, B, C, and D, are depicted with corresponding descriptions on the left-hand side of the figure.  (It is also assumed that correlations between separate source points -- such as that manifest in the Pfleegor-Mandel effect involving coherent radiation from a pair of identical but independent lasers -- does not occur or is negligible [21]).*

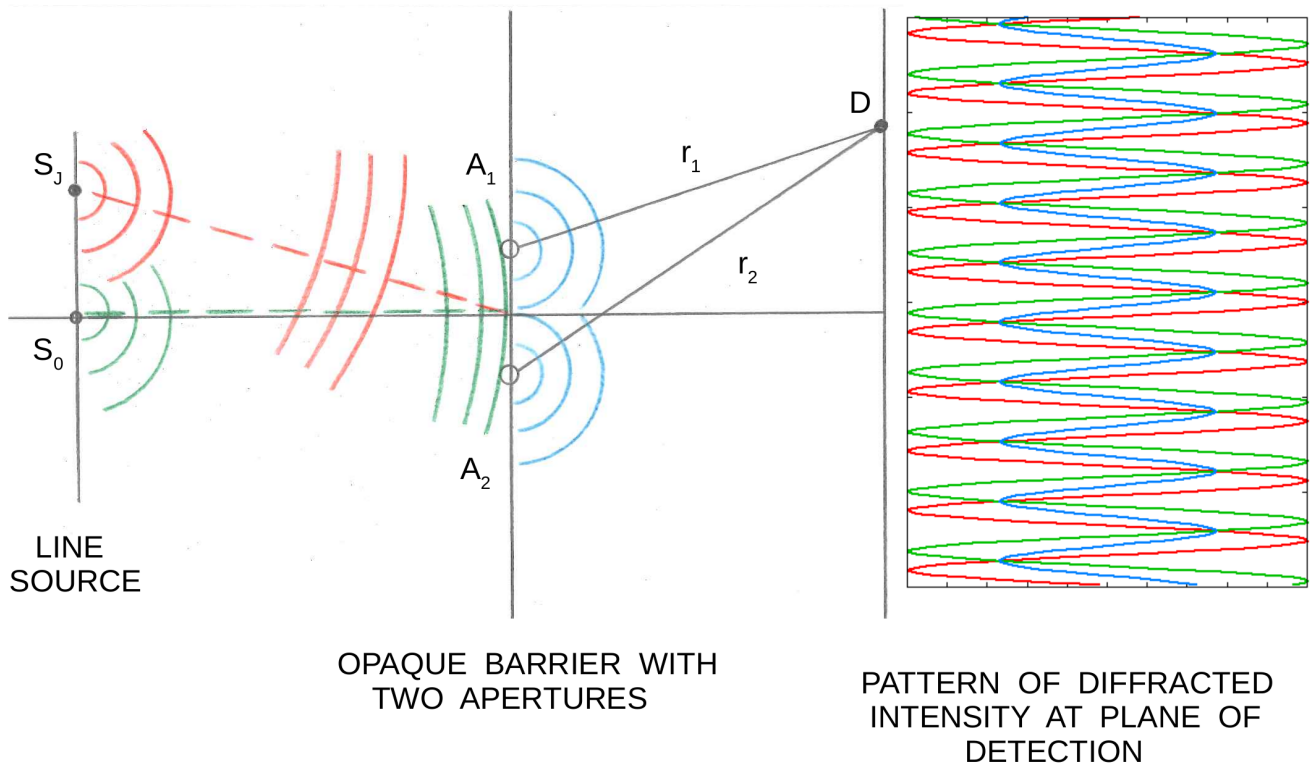

LINE
SOURCE

OPAQUE BARRIER WITH
TWO APERTURES

PATTERN OF DIFFRACTED
INTENSITY AT PLANE OF
DETECTION

*Figure 8. Young's experiment for creating an interference pattern in which two apertures are illuminated with quasi-monochromatic radiation from a temporally and spatially extended incoherent source. In this two-dimensional illustration, the radiation is emitted as circular waves from each point on a line. The distances between source and apertures and between apertures and detection line are great enough that the Fraunhofer or far-field limit is a valid approximation for describing the scattering. At the location of the apertures, the circular wave fronts are nearly planar, as described in the text for the purposes of the analysis performed. Any single-aperture diffraction that might occur and modulate the two-slit interference pattern plotted on the far right-hand side of the figure is neglected since it is not relevant to the arguments made concerning the role of the mutual coherence function and fringe visibility discussed in the text. The red and green patterns of intensity plotted on the right correspond to two point sources, one at the origin $S_0$ and the other off the horizontal axis of symmetry at $S$, respectively. The blue curve results upon adding theses two single-source-point intensity distributions together. Because of the translational offset between the two single-point patterns, a reduced "fringe" visibility or diminished instrumental resolution occurs.*

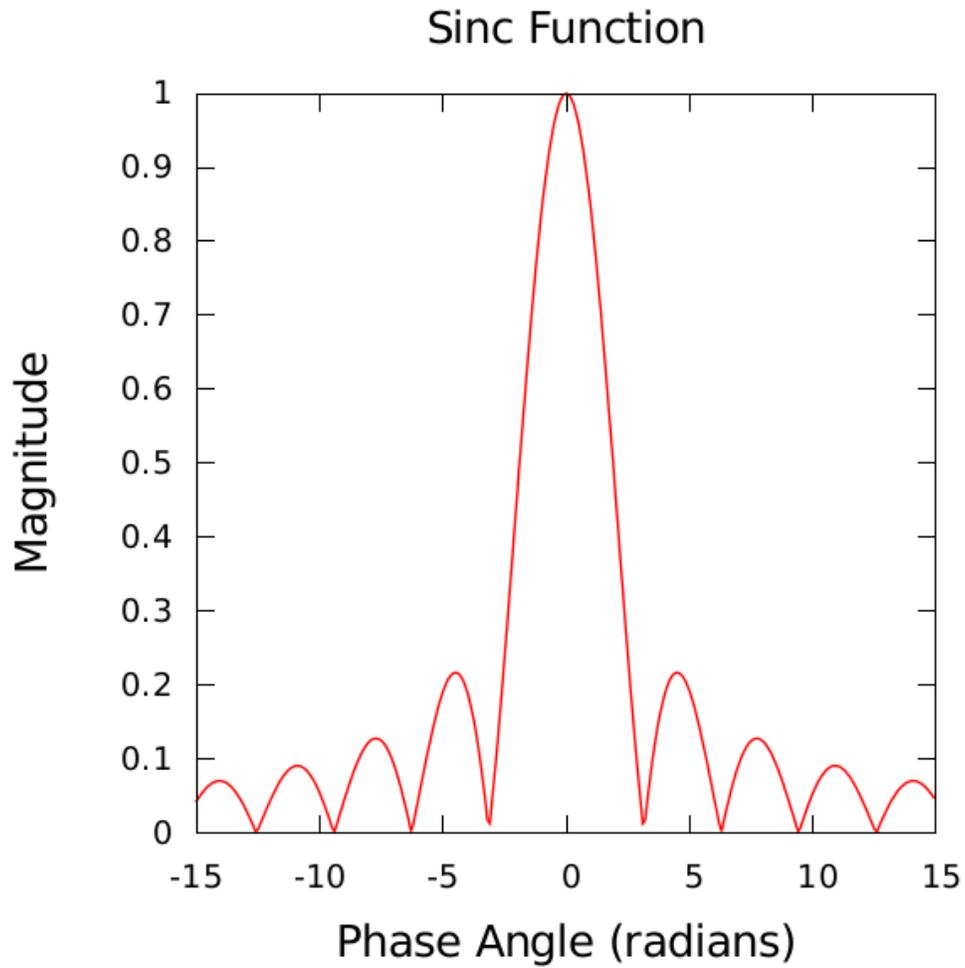

Figure 9. Plot of the modulus of the complex degree of coherence function $\gamma_{A1A2}$ for the case of quasi-monochromatic radiation from a temporally and spatially extended incoherent source for which the wave fronts are nearly planar at the positions of points $A_1$ and $A_2$ as described in the text.

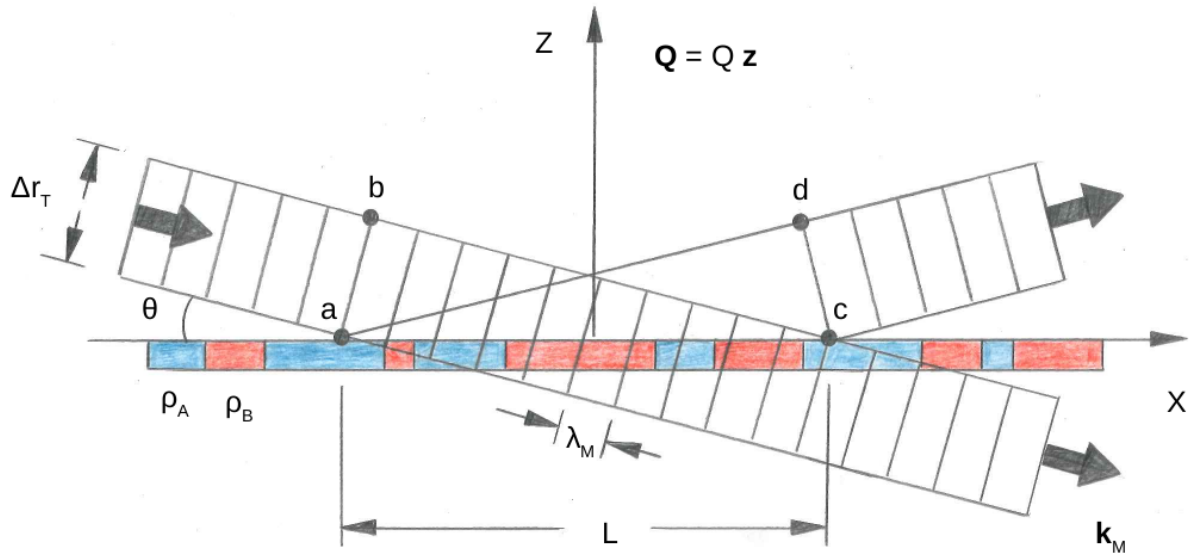

*Figure 10. Schematic representation of a wave packet of rectangular form interacting with a planar sample of inhomogeneous SLD in plane (two values, $\rho_A$ and $\rho_B$). This picture illustrates how, for elastic, coherent, specular scattering, the area of the scattering surface that a wavefront of constant phase "sees" along the horizontal axis in the figure is actually its transverse dimension $\Delta r_T$ projected a length L across the surface. The other, orthogonal width (along an axis perpendicular to the plane of the figure itself) in the plane seen by the wavefront is not amplified but equal to whatever the packet width is in that direction. The lower edge of the jth wavefront intersects the sample surface first, on the left, and then the upper edge a distance L farther along. Note that the distances a to d and b to c are equal -- applying the Hugens-Fresnel wavelet construction shows that in the specular condition the incident planar wavefront ab is exactly in phase with the reflected wavefront cd (assuming a perfectly flat material reflecting surface).*

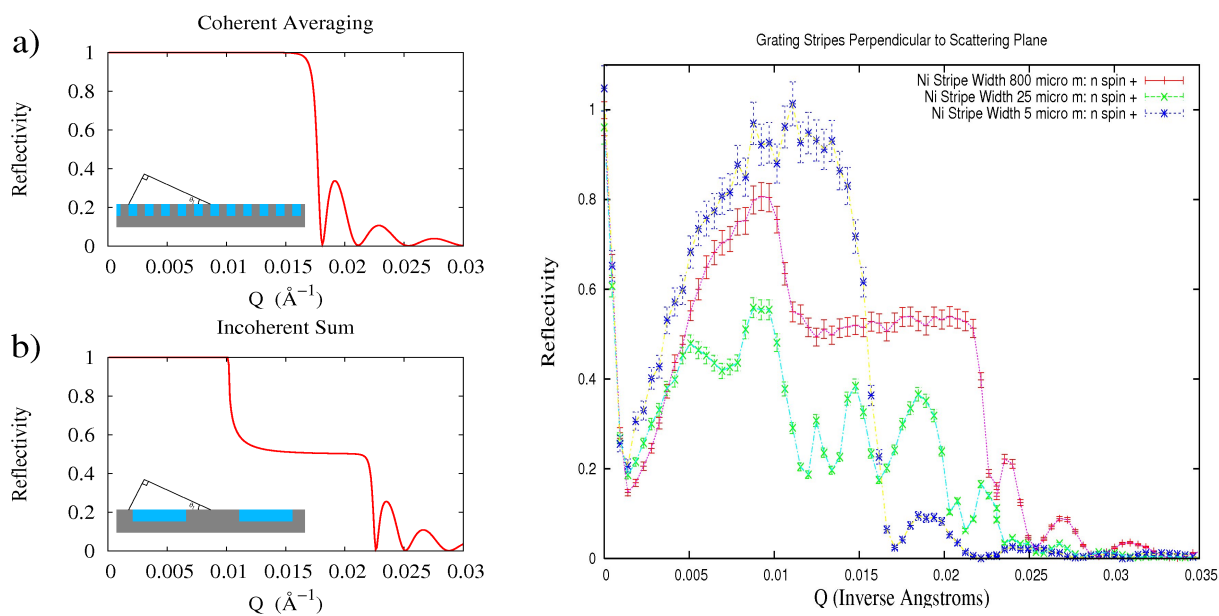

Figure 11. *Summary of one of the principal results of previous work [4] in which it was originally demonstrated that the specular reflectivity about the critical edge for external mirror reflection is a sensitive measure of the projected length of a neutron wavefront (after Figures 7 and 15 of [4]). (a) Model specular reflectivity curve corresponding to an effective coherent averaging of two different SLDs. In the real space schematic of the grating structure in the inset, the material for the periodic rectangular grating structure is the same as that of the substrate and is taken to have the SLD of Si; the troughs in between, on the other hand, are filled with material having the SLD of ordinary Ni (bar and trough widths are equal). Only a single critical Q is observed. (b) Model specular reflectivity curve corresponding to an incoherent sum of two independently scattering areas of in-plane SLD in the film. Two distinct critical Q values appear in this case. Both of the model reflectivity curves plotted in (a) and (b) were calculated for the case of perfect instrumental resolution -- i.e., a monochromatic beam with no angular divergence. On the right hand side of the figure are shown experimental specular reflectivity data for the two limiting cases (in addition to an intermediate case) [4].*

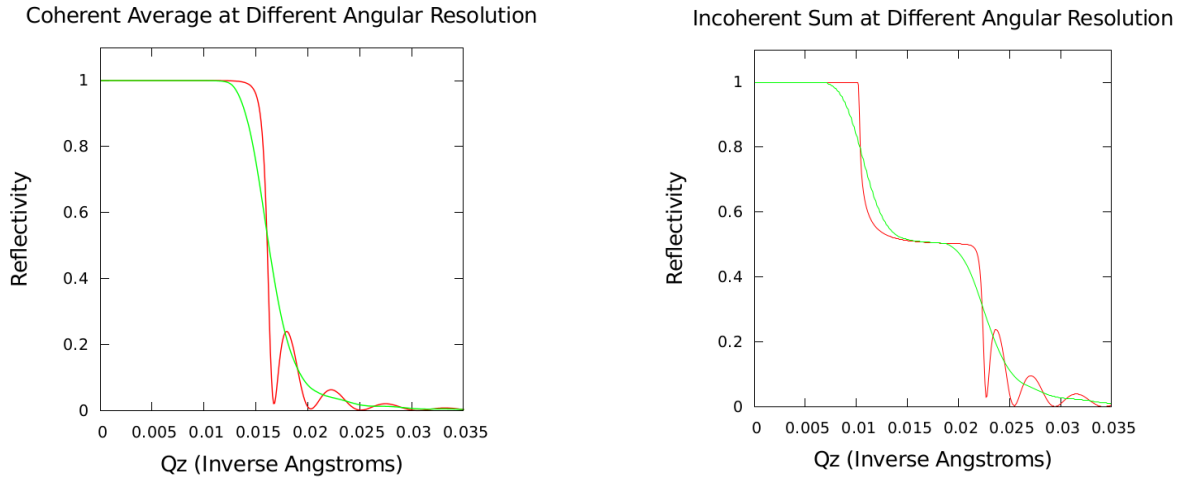

*Figure 12. Model calculations of the specular neutron reflectivity as a function of Q for different beam angular resolutions in the two limiting cases. In the left-hand plot the transverse dimension of the wavefront is of sufficient extent to completely average over a large enough number of the bars and troughs of the grating structure. Conversely, in the right-hand plot the transverse dimension of the wavefront was significantly less than the width of a bar or trough (bar width = trough width). The substrate was taken to be silicon with approximately 950 Angstrom thick nickel bars deposited on top. Neutrons were taken to be incident from vacuum. Both plots show specular reflectivity curves at two extremes of instrumental angular beam divergence, approximately 3.5 x 10⁻⁵ and 1.5 x 10⁻³ radians, at a fractional wavelength resolution of 0.01. Despite a difference of a factor of over 40 in the angular divergence of the beam and the consequential rounding of the critical edge and smearing of the film thickness fringes at the broader angular divergences, an unambiguous distinction between the cases for coherent averaging and incoherent sum can be made.*

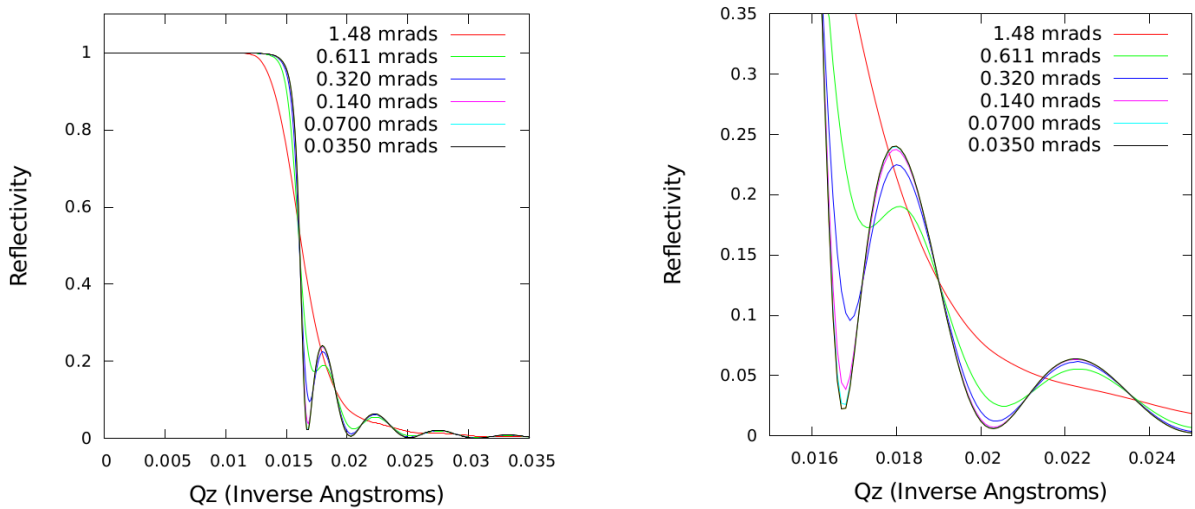

*Figure 13. (Left) Model calculation showing a reduction in the Kiessig fringe visibility with broadening beam angular divergence as predicted for the instrumental $Q_z$ - resolution perpendicular to the grating film surface. (Right) Detail about the first two fringes. The geometrical beam angular divergence ranges from 3.5 x 10⁻⁵ to 1.48 x 10⁻³ radians.*

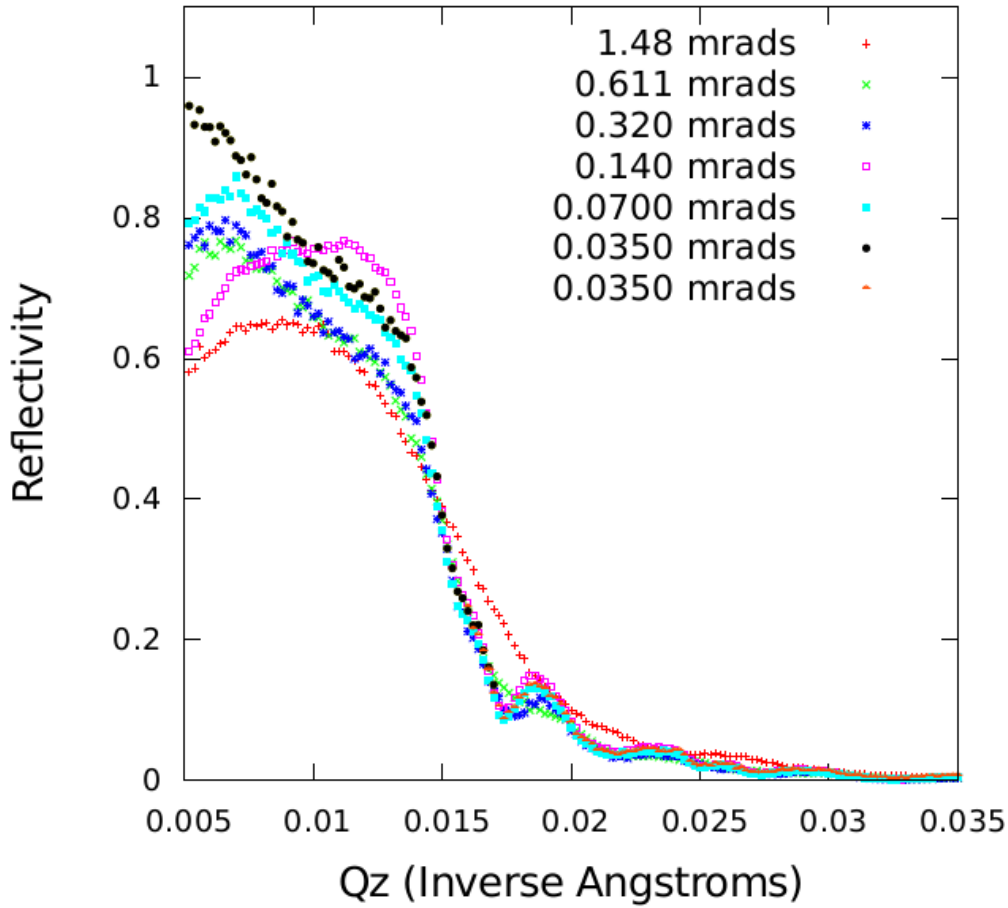

*Figure 14. Composite plot of the measured specular reflectivites as a function of beam angular divergence for the 10 + 10 = 20 micron grating with the mean beam wavevector perpendicular to the grating bars. At all angular beam divergences, the individual neutron packets effectively average over bars and troughs resulting in a single critical cutoff corresponding to the average SLD of the Ni bars and the empty spaces in between. The values of the angular beam divergence and the corresponding aperture widths are given in Table 4. The specular reflectivity (reflected intensity divided by that incident) for a homogeneous film layer should plateau at nearly unit reflectivity below the critical angle -- however, for the grating structure here it dips to about 85 % approaching the critical $Q_C$ because of competing non-specular scattering from the periodic (but inhomogeneous) in-plane grating structure. The presence of non-specular scattering was confirmed in scans along the $Q_X$ axis as well as in detector two-theta or scattering angle scans at fixed sample or theta angles. This effect is essentially irrelevant to the specular measurements regarding either beam angular resolution or transverse packet dimensions. The slight downturn in reflected intensity below $Q \approx 0.0075$ $Å^{-1}$ approaching the origin is due to the substrate not intercepting the entire footprint of the beam width (which would require an infinitely long substrate at zero). Although the angular divergence of the incident beam was varied by more than a factor of 40 (see Table 4), the neutron wave packet had a transverse extent sufficient to average over a significant number of Ni stripes and intervening troughs thereby resulting in a single critical edge at the mean value of SLD.*

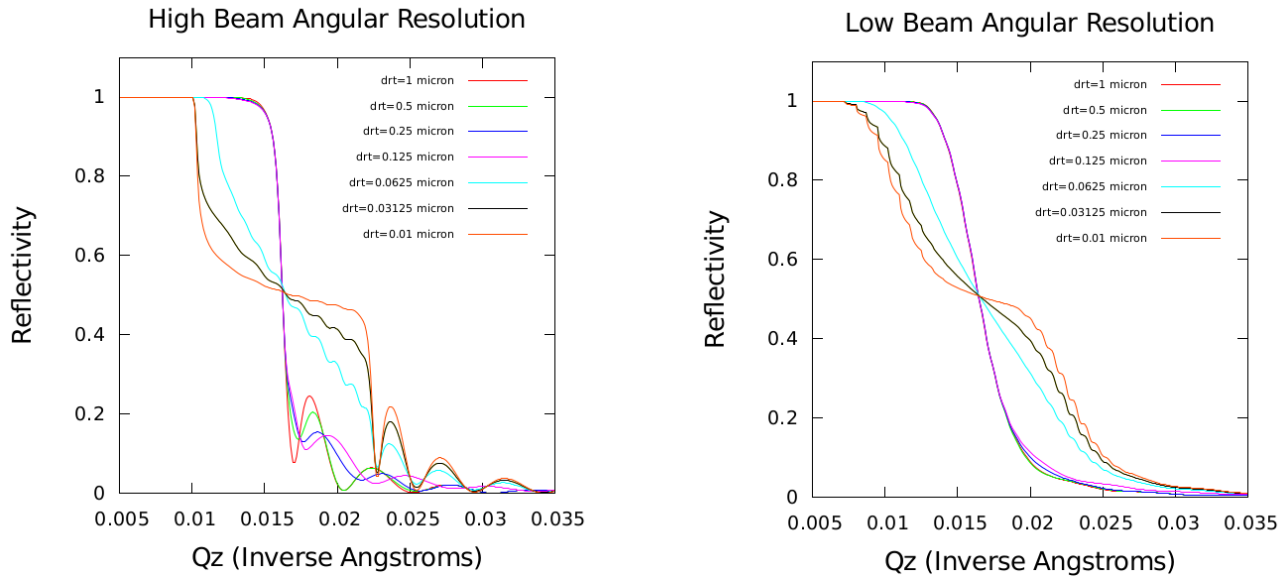

*Figure 15. Calculated model specular neutron reflectivity curves about the effective critical angle for a 10 + 10 = 20 micron grating (Ni stripes 950 Angstroms thick) with neutron wavevector perpendicular to the Ni stripes at relatively high (7.0 x 10⁻⁵ radians, left plot) and low (1.5 x 10⁻³ radians, right plot) instrumental beam angular resolution -- for different values of $\Delta r_T$ where the glancing angular dependence of the projection given by Equation 16, i.e., $\Delta r_T = L \sin \theta$, was explicitly taken into account. Despite the marked difference in instrumental angular beam divergence, the transverse extent of uniform phase of a neutron packet wavefront has a clearly distinguishable effect on the specular reflectivity in the critical angle region in both cases.*

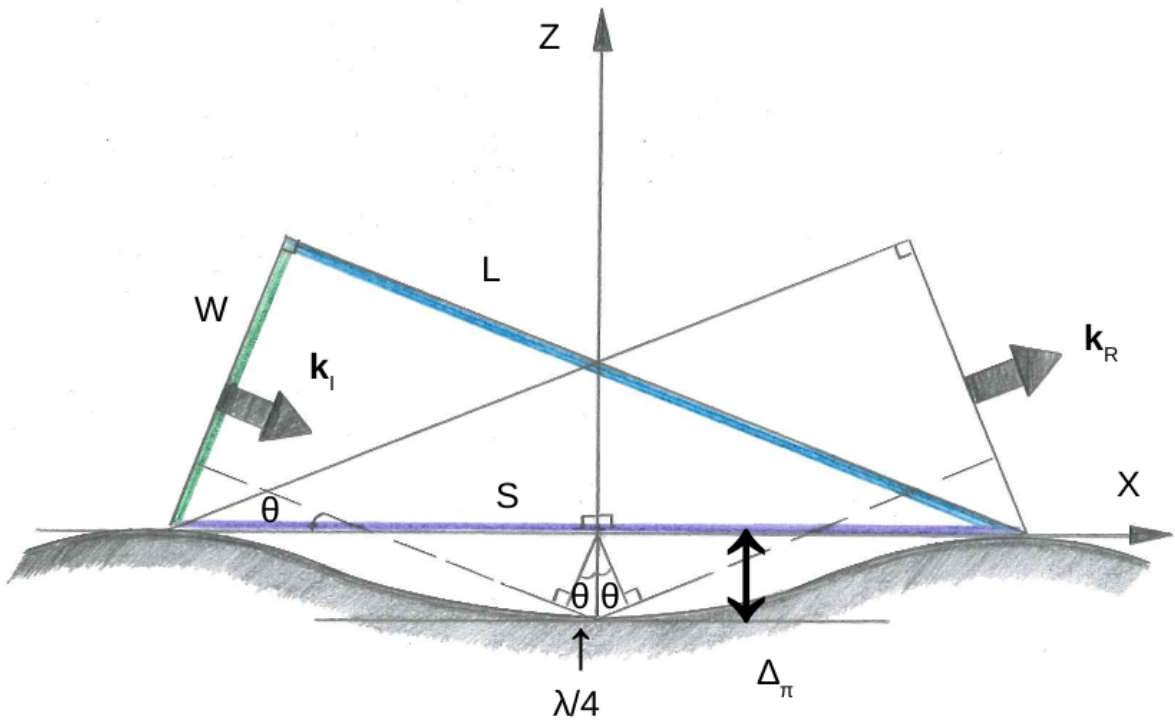

*Figure 16. Simple representation of a wavy surface as a sinusoid. The wavy surface causes a phase shift between different regions of the incident and specularly reflected wavefronts as described in the text. W, L, and S label certain distances used in the derivation of the measure of deviation of a curved surface from perfect flatness discussed in the text.*

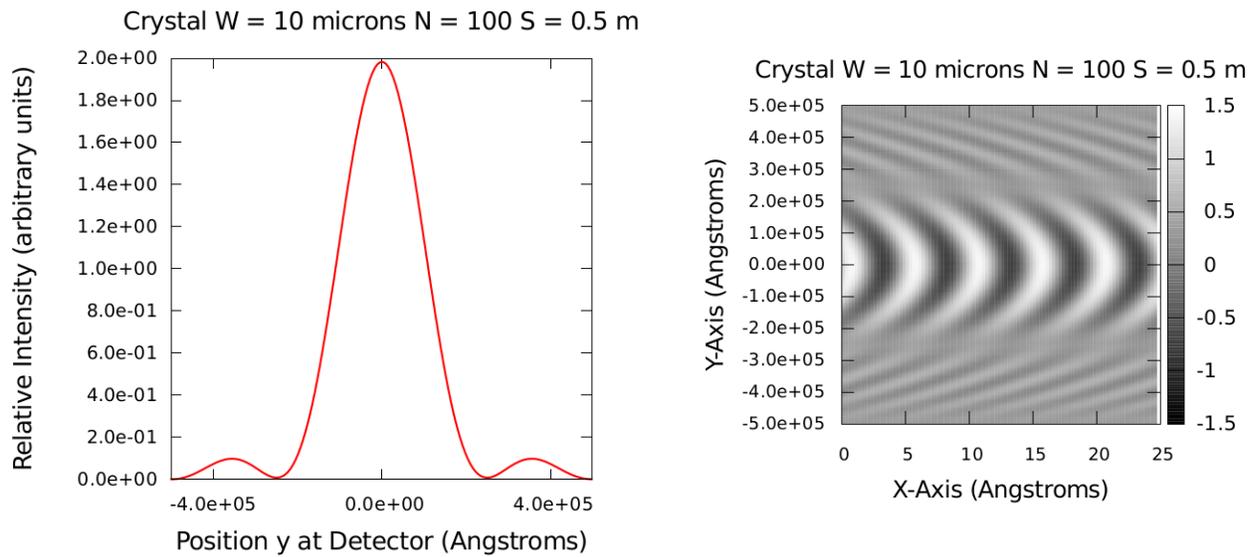

*Figure B 1. Results of a Huygens-Fresnel calculation assuming a generic crystal 10 microns wide with 100 reflecting atomic planes spaced of 5 Å apart from each other, and with 1000 atomic source points per plane. The wavelength was taken to be 5 Å as well and the distance between crystal face and point of observation of the reflected wave was 0.5 meter. This two-dimensional block of source points was taken to be illuminated by plane waves in phase as though the Bragg diffraction condition was effectively satisfied. As in the case of an aperture of the same width, the reflected wave has a well-defined lateral dimension which at 0.5 m from its source has a uniform wavefront (to within one wavelength) over a lateral extent of approximately 22. microns -- similar to that produced by the single aperture of the same width. On the left is a plot of intensity versus position on a perpendicular detection plane a distance 0.5 m away from the crystal. The horizontal axis of the wave amplitude plot on the right is along the mean direction of propagation covering a distance of approximately five wavelengths up to the detection plane at 0.5 m at the right terminus -- the vertical axis is along a perpendicular direction and the degree of shading indicates the relative amplitude of the reflected wave.*

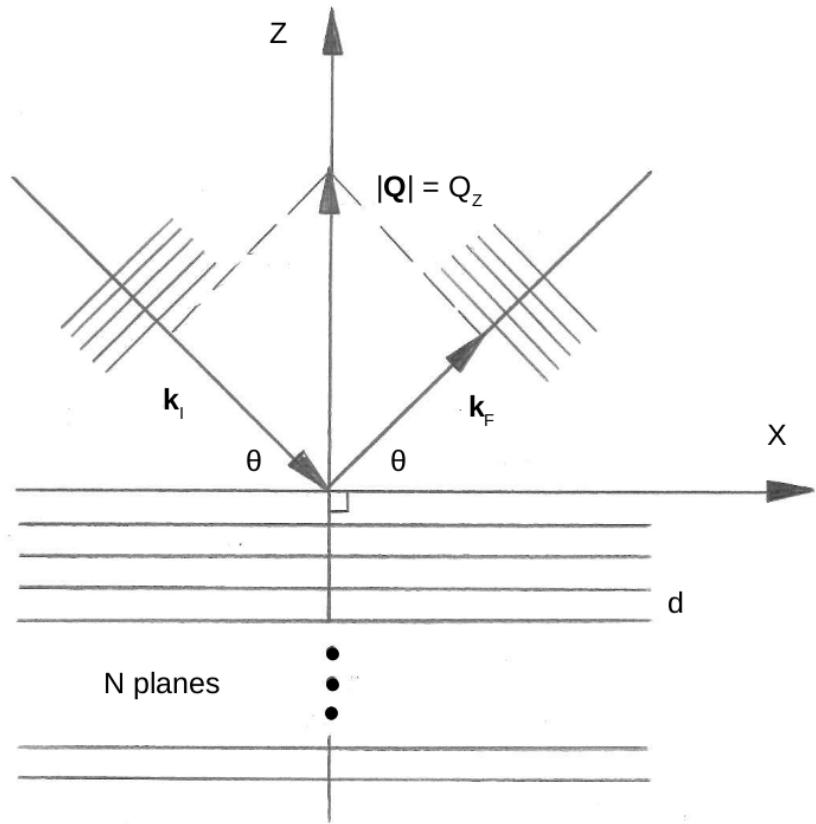

Figure B 2. *Schematic of the diffraction geometry from a perfect single mosaic block consisting of a set of N parallel atomic planes spaced a distance d apart from one another. The incident and scattered wavevectors are $k_I$ and $k_F$, respectively, while $Q = k_F - k_I$ is the wavevector transfer ($|k_I| = |k_F| = k$ since the scattering is elastic). The specular condition is shown in which the angles of incidence and reflection relative to the surface are equal and denoted as $\theta$.*

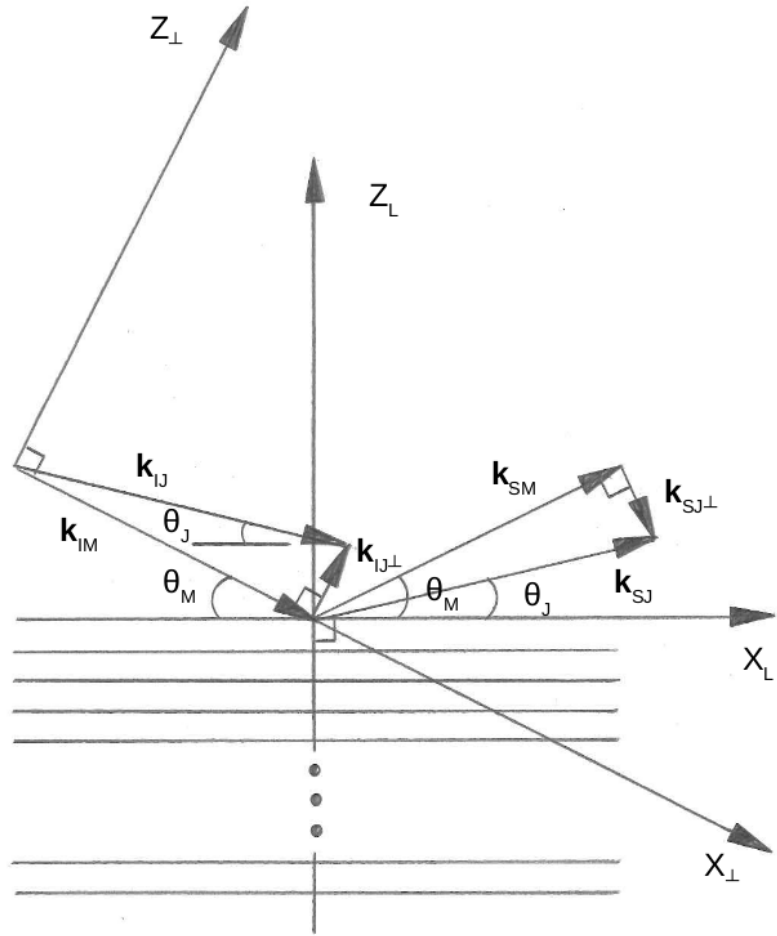

*Figure B 3. Schematic of the reflection geometry. In this diagram, the reference frame of the incident wave is shown in addition to the laboratory or crystalline block reference frame. The $z_\perp$ and $x_\parallel$ orthogonal axes define the neutron frame whereas $z_L$ and $x_L$ correspond to the laboratory frame. $\mathbf{k}_{Ij}$ labels the jth basis wavevector component of the incident wave train distribution yhat is incident at an angle $\theta_j$ relative to the surface atomic plane of the crystal. $\mathbf{k}_{IM}$, on the other hand, labels the mean or central wavevector of the wave train. The corresponding scattered or reflected wavevectors are labeled with the subscript "s" in an analogous way.*

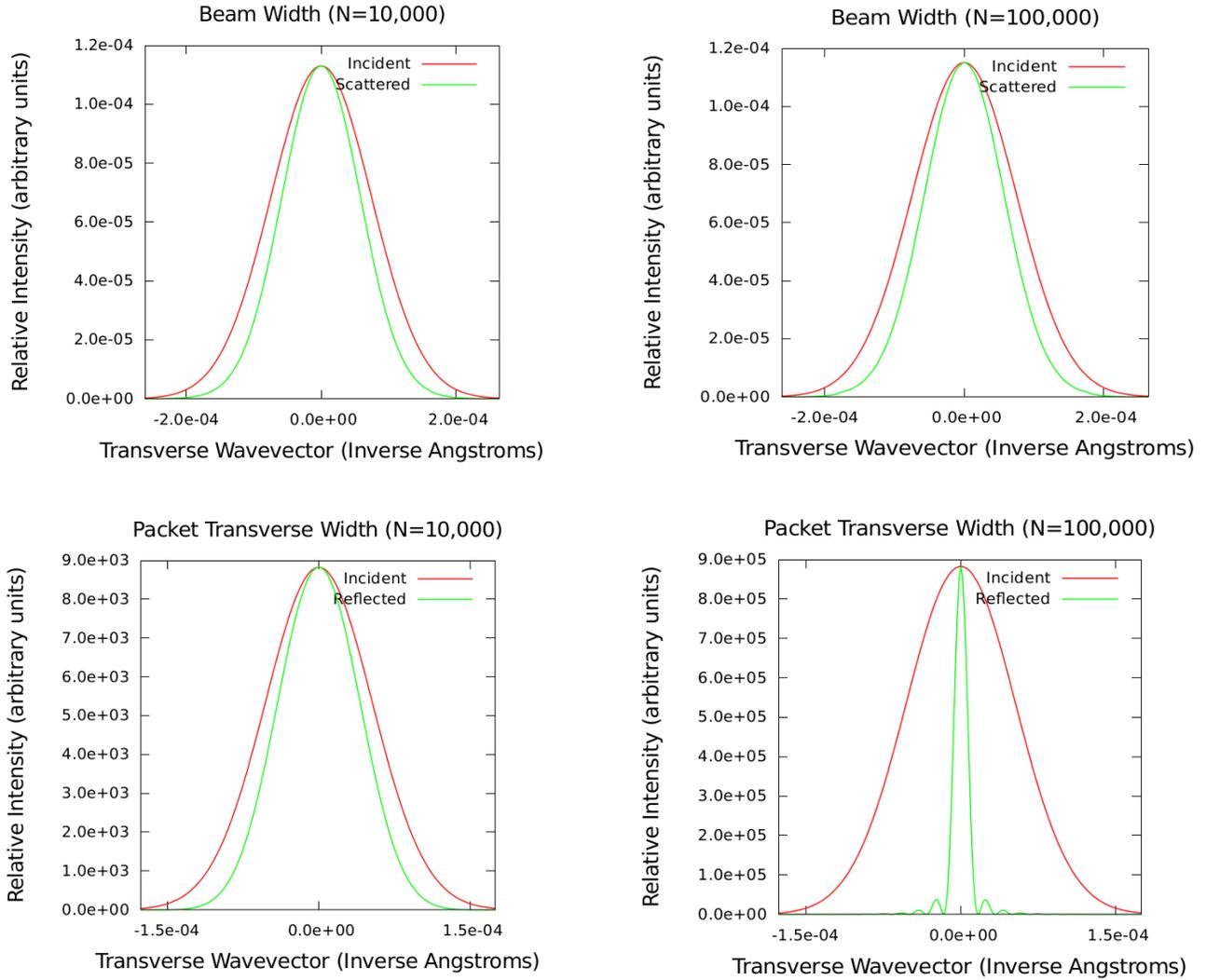

*Figure B 4. Model reflectivity curves for a Si(111) perfect single crystal block of finite size calculated according to the approximate theory introduced in the text. The examples correspond to crystal blocks with N equal to either $10^4$ (left) or $10^5$ (right) atomic planes. An incident beam divergence of 1.4 x $10^{-4}$ radians (FWHM) and an incident packet transverse wavevector component distribution width $\Gamma_{FWHM\,k\perp}$ ( = $\Delta k_T$) of 1.76 x $10^{-4}$ Å$^{-1}$ are assumed.*

**(Upper Two Plots) -- Beam Width**  *Examples of the incident and reflected beam intensities as a function of the transverse component of the packet mean wavevector $k_{MT} = k_M \tan(\alpha)$ -- where $\alpha$ is the angle of deviation a given neutron packet's mean wavevector $k_M$ makes relative to the average packet mean wavevector $<k_M>$ in a geometrically collimated beam.*

**(Lower Two Plots) -- Packet Transverse Width**  *Although the Bragg diffraction process narrows the angular divergence of the reflected beam, it also simultaneously reduces the width of the transverse wavevector component distribution intrinsic to each individual reflected wave packet. As discussed in the text, the narrowing is inversely proportional to the number of atomic planes contributing to the coherent reflection process. The narrowing of the reflected packet transverse wavevector component distribution becomes more pronounced with increasing width of the incident packet's distribution. If the incident distribution is already relatively narrow, the width of the reflected distribution may not be significantly reduced further.*

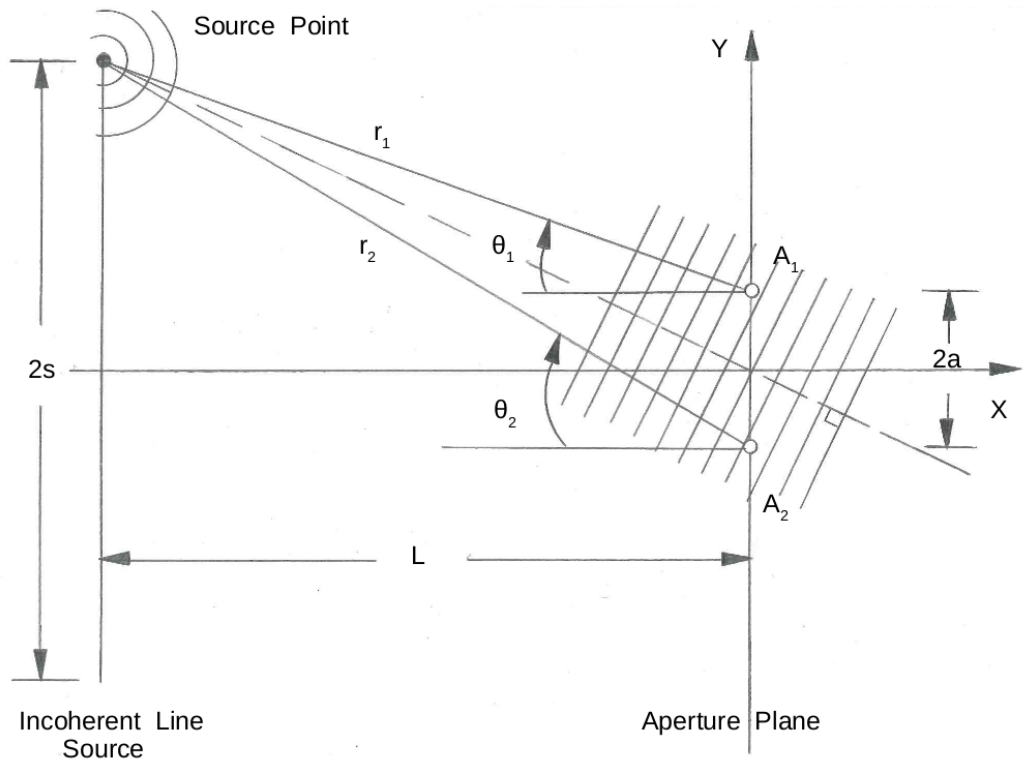

*Figure C1.  The relevant part of Figure 8 is redrawn in more detail here for reference to the discussion in the text regarding the calculation of the mutual coherence function.*

**Diffraction of an Incident Plane Wave of 5 Å Wavelength by a 2D Slit Aperture**

| W (Å) | S (Å) | $\Gamma_{FWHM}$ ( Å) (Intensity Distribution) | $\Gamma_{FWHM}$ ( Å) | $\Delta r_T$ (Å) | $\Delta\theta$ (rad) |
|---|---|---|---|---|---|
| Slit Width | Slit to Detector Distance | via Huygens-Fresnel (HF) Construction | Fraunhofer Limit (Analytic) | H-F Amplitude | $\sim \Gamma/S$ |
| $10^4$ (1 μm) | $0.5 \times 10^{10}$ | $2.48 \times 10^6$ | $2.50 \times 10^6$ | $2.19 \times 10^5$ (21.9 μm) | $5.0 \times 10^{-4}$ |
| $10^5$ (10 μm) | $0.5 \times 10^{10}$ | $2.53 \times 10^5$ | $2.50 \times 10^5$ | $2.18 \times 10^5$ (21.8 μm) | $5.0 \times 10^{-5}$ |
| $10^6$ (100 μm) | $0.5 \times 10^{10}$ | $8.92 \times 10^5$ | ---------- | $9.26 \times 10^5$ (92.6 μm) | $1.8 \times 10^{-4}$ |
| $10^5$ (10 μm) | $2.5 \times 10^{10}$ | $1.25 \times 10^6$ | $1.25 \times 10^6$ | $4.63 \times 10^5$ (46.3 μm) | $5.0 \times 10^{-5}$ |
| $10^6$ (100 μm) | $2.5 \times 10^{10}$ | $7.74 \times 10^5$ | ---------- | $7.41 \times 10^5$ (74.1 μm) | $3.1 \times 10^{-5}$ |

(For Fraunhofer limit, analytic $y_{MIN} \approx$ S λ/W which is *approximately* $\Gamma_{FWHM}$ .)

*Table 1.   Results of a Huygens-Fresnel construction for a variety of pertinent aperture widths and distances to a plane of observation (see Figure 2 for schematic).  Plots of the real part of the reflection amplitude as well as the intensity distribution at a plane of observation a distance S from the aperture are plotted in Figure 3 for two of the cases listed in the table.*

| Incident Beam Angular Divergence | | | | $\Delta k_{MT(BEAM)}$ | N | $\Delta r_{TWP}$ | $\Delta k_{TWP}$ |
| (rad) | (deg) | (min) | (sec) | (Å⁻¹) | | (μm) | (Å⁻¹) |
| --- | --- | --- | --- | --- | --- | --- | --- |
| $1.75 \times 10^{-5}$ | 0.001 | 0.06 | 3.6 | $2.19 \times 10^{-5}$ | 2 | 4.8 | $1.04 \times 10^{-5}$ |
| $3.50 \times 10^{-5}$ | 0.002 | 0.12 | 7.2 | $4.38 \times 10^{-5}$ | " | " | " |
| $1.75 \times 10^{-5}$ | 0.001 | 0.06 | 3.6 | $2.19 \times 10^{-5}$ | 4 | 9.6 | $5.20 \times 10^{-6}$ |
| $3.50 \times 10^{-5}$ | 0.002 | 0.12 | 7.2 | $4.38 \times 10^{-5}$ | " | " | " |
| $1.75 \times 10^{-5}$ | 0.001 | 0.06 | 3.6 | $2.19 \times 10^{-5}$ | 30 | 72.0 | $6.90 \times 10^{-7}$ |
| $3.50 \times 10^{-5}$ | 0.002 | 0.12 | 7.2 | $4.38 \times 10^{-5}$ | " | " | " |

where $\Delta r_{TWP} \Delta k_{TWP} = 1/2$ or $\Delta k_{TWP} = 1 / (2 \Delta r_{TWP})$ and $\Delta k_{MT(BEAM)} = \sin(\Delta \alpha)$ where $\Delta \alpha$ is the FWHM of the angular divergence distibution of the incident beam.

*Table 2. Comparison of the differences between the magnitudes of the widths of the transverse wavevector component distributions associated with an individual wave packet and that of the packet mean wavevectors contained in the beam. For a beam angular divergence $\Delta \alpha$ = 7.2 arc seconds, the ratios of the FWHM of the packet mean wavevector distribution $\Delta k_{MT(BEAM)}$ divided by an individual packet's component basis wavevector distribution $\Delta k_{TWP}$ are approximately 4, 8, and 63 for N = 2, 4, and 30, respectively. (The nominal neutron wavelength is 5 Å.)*

Table 3. Values of the Critical Wavevector $Q_C$ for Relevant Materials -- $Q_C^2 = 16 \pi \rho$

| Material | $Q_C$ ( Å⁻¹ ) |
| --- | --- |
| Ni (unmagnetized) | 0.0217 |
| Si | 0.0102 |
| 50 % Ni + 50 % Si (by volume) | 0.0170 |
| 50 % Ni + 50 % vacuum (by volume) | 0.0154 |

# Table of Aperture Widths, Angular Beam Divergence, and Instrumental Resolution

| $W_1$ | $W_2$ (mm) | $L_{12}$ | $\Delta\theta_{BM}$ (radians) | $2\Delta k_{BMT} \approx 2k_{BM}\,\Delta\theta_{BM}$ (inverse Angstroms) | $1/2\Delta k_{BMT}$ (μm) |
|---|---|---|---|---|---|
| 0.05 | 0.05 | 1429. | $3.50 \times 10^{-5}$ | $8.80 \times 10^{-5}$ | 1.136 |
| 0.10 | 0.10 | 1429. | $7.00 \times 10^{-5}$ | $1.76 \times 10^{-4}$ | 0.568 |
| 0.20 | 0.20 | 1429. | $1.40 \times 10^{-4}$ | $3.52 \times 10^{-4}$ | 0.284 |
| 1.00 | 0.10 | 1719. | $3.20 \times 10^{-4}$ | $8.04 \times 10^{-4}$ | 0.124 |
| 2.00 | 0.10 | 1719. | $6.11 \times 10^{-4}$ | $1.54 \times 10^{-3}$ | 0.0649 |
| 5.00 | 0.10 | 1719. | $1.48 \times 10^{-3}$ | $3.72 \times 10^{-3}$ | 0.0269 |

*Table 4. Typical reflectometer slit widths and geometrical angular divergences for the incident beam corresponding to the data shown in Figure 14. The angular widths (FWHM) calculated from the slit widths and their separation distance are typically measured to be consistent to within a few (2 to 3) percent accuracy.*